\documentclass{emulateapj}
\usepackage{graphicx}

\newcommand{\mum}{\ifmmode{\rm \mu m}\else{$\mu$m}\fi}
\newcommand{\iso}{{\em ISO}}

\begin{document}

\title{The infrared spectral properties of Magellanic carbon stars}

\author{
G.~C.~Sloan\altaffilmark{1,2},
K.~E.~Kraemer\altaffilmark{3},
I.~McDonald\altaffilmark{4},
M.~A.~T.~Groenewegen\altaffilmark{5},
P.~R.~Wood\altaffilmark{6},
A.~A.~Zijlstra\altaffilmark{4},
E.~Lagadec\altaffilmark{7},
M.~L.~Boyer\altaffilmark{8,9},
F.~Kemper\altaffilmark{10},
M.~Matsuura\altaffilmark{11},
R.~Sahai\altaffilmark{12},
B.~A.~Sargent\altaffilmark{13},
S.~Srinivasan\altaffilmark{10},
J.~Th.~van Loon\altaffilmark{14},
\& K.~Volk\altaffilmark{15}
}
\altaffiltext{1}{Cornell Center for Astrophysics \& Planetary Science, 
  Cornell Univ., Ithaca, NY 14853-6801, USA, sloan@isc.astro.cornell.edu}
\altaffiltext{2}{Department of Physics and Astronomy, University of North
  Carolina, Chapel Hill, NC 27599-3255, USA}
\altaffiltext{3}{Institute for Scientific Research, Boston College,
  140 Commonwealth Avenue, Chestnut Hill, MA 02467, USA}
\altaffiltext{4}{Jodrell Bank Centre for Astrophysics, Univ.\ of Manchester, 
  Manchester M13 9PL, UK}
\altaffiltext{5}{Koninklijke Sterrenwacht van Belgi\"e, Ringlaan 3, B--1180 
  Brussels, Belgium}
\altaffiltext{6}{Research School of Astronomy and Astrophysics, Australian
  National University, Canberra, ACT 2611, Australia}
\altaffiltext{7}{Observatoire de la C\^{o}te d'Azur, 06300, Nice, France}
\altaffiltext{8}{CRESST and Observational Cosmology Lab, Code 665, NASA 
  Goddard Space Flight Center, Greenbelt, MD, 20771, USA}
\altaffiltext{9}{Department of Astronomy, University of Maryland, College 
  Park, MD 20742, USA}
\altaffiltext{10}{Academia Sinica, Institute of Astronomy and Astrophysics,
  11F Astronomy-Mathematics Building, NTU/AS, No. 1, Sec. 4, Roosevelt Rd.,
  Taipei 10617, Taiwan, R.O.C.}
\altaffiltext{11}{School of Physics and Astronomy, Cardiff University, Queen's 
  Buildings, The Parade, Cardiff, CF24 3AA, UK}
\altaffiltext{12}{Jet Propulsion Laboratory, California Institute of 
  Technology, MS 183-900, Pasadena, CA 91109, USA}
\altaffiltext{13}{Center for Imaging Science and Laboratory for 
  Multiwavelength Astrophysics, Rochester Institute of Technology, 54 Lomb 
  Memorial Drive, Rochester, NY 14623, USA}
\altaffiltext{14}{Lennard Jones Laboratories, Keele University, Staffordshire
  ST5 5BG, UK}
\altaffiltext{15}{Space Telescope Science Institute, 3700 San Martin Dr., 
  Baltimore, MD 21218, USA}

\begin{abstract}
The Infrared Spectrograph on the {\it Spitzer Space 
Telescope} observed 184 carbon stars in the Magellanic 
Clouds.  This sample reveals that the dust-production rate 
(DPR) from carbon stars generally increases with the 
pulsation period of the star.  The composition of the dust 
grains follows two condensation sequences, with more SiC 
condensing before amorphous carbon in metal-rich stars, and 
the order reversed in metal-poor stars.  MgS dust condenses 
in optically thicker dust shells, and its condensation is 
delayed in more metal-poor stars.  Metal-poor carbon stars 
also tend to have stronger absorption from C$_2$H$_2$ at 
7.5~\mum.  The relation between DPR and pulsation period 
shows significant apparent scatter, which results from the 
initial mass of the star, with more massive stars occupying a 
sequence parallel to lower-mass stars, but shifted to longer 
periods.  Accounting for differences in the mass distribution 
between the carbon stars observed in the Small and Large 
Magellanic Clouds reveals a hint of a subtle decrease in the 
DPR at lower metallicities, but it is not statistically 
significant.  The most deeply embedded carbon stars have
lower variability amplitudes and show SiC in absorption.  
In some cases they have bluer colors at shorter wavelengths,
suggesting that the central star is becoming visible.
These deeply embedded stars may be evolving off of the AGB 
and/or they may have non-spherical dust geometries.
\end{abstract}


\section{Introduction} % Sec. 1.0

Intermediate-mass stars contribute to the chemical evolution 
of galaxies when they shed their envelopes and seed their 
environments with freshly condensed dust and the products of 
nuclear fusion from their cores.  The sensitivity of the 
Infrared Spectrograph \citep[IRS;][]{hou04} on the {\it 
Spitzer Space Telescope} \citep{wer04} made it possible to 
observe many mass-losing stars in nearby galaxies like the 
Large and Small Magellanic Clouds (LMC and SMC).  These stars 
are at known distances,
and they probe different metallicities.  While the metallicity 
distribution functions are broad enough to overlap and depend 
on age, for statistical treatment of large samples it is 
safe to claim that the LMC is more metal-poor than the 
Galaxy, and the SMC is more metal-poor still.  As an example,
\cite{pia13} find that in the LMC, [Fe/H] increased from
$\sim$$-$0.7 to $\sim$$-$0.3 from 5 to 1 Gyr ago, compared to 
$\sim$$-$1.2 to $\sim$$-$0.6 over the same time interval in
the SMC \citep{pia12}.
%\citep[see, for example, the age-metallicity relations by][]{pia12,pia13}.
% SMC:  1 Gyr, [Fe/H] ~ -0.6, 5 Gyr, ~ -1.2 (pia12)
% LMC:  1 Gyr, [Fe/H] ~ -0.3, 5 Gyr, ~ -0.7 (pia13)

Carbon stars dominated the spectral samples of evolved 
stars observed early in the {\it Spitzer} mission.
\cite{slo06} examined carbon stars in the SMC while 
\cite{zij06} considered the LMC.  \cite{mat06} studied the 
molecular absorption in these samples, and other studies 
followed \citep{buc06, lag07, lei08, slo08}.  
Further observing programs added considerably to 
the spectral sample of carbon stars in both galaxies,
including two programs which added fainter carbon stars
to the samples in the LMC \citep{kem10} and SMC (Program
50240).

These earlier efforts led to the finding that the amount of 
dust observed around carbon stars does not show a strong
dependence on metallicity \citep{slo08}.  Carbon stars result 
from the dredge-up of freshly fused carbon from the cores of 
stars as they burn He in thermal pulses while on the 
asymptotic giant branch (AGB).  The formation of CO leaves 
only carbon or oxygen available to form other molecules and 
condense into dust so that when the C/O ratio exceeds unity, 
the gas and dust shift from an O-rich to a C-rich chemistry.
\cite{mat05} noted that because dredged-up material on the
AGB is dominated by carbon in most stars, the amount of free 
carbon left after the formation of CO should actually {\it 
increase} in more metal-poor stars.  \cite{slo12} accounted
for the changing oxygen abundances at lower metallicity and
reached a similar conclusion.  Thus, the question is not one 
of why the amount of carbon-rich dust is not {\it declining} 
as metallicity drops, but why is it not {\it increasing}.

The infrared spectrographic studies of Magellanic carbon
stars have revealed several dependencies on metallicity.  
In more metal-poor samples, the SiC dust emission feature at 
$\sim$11.3~\mum\ weakens, the molecular absorption bands 
from C$_2$H$_2$ at 7.5 and 13.7~\mum\ grow stronger, and the
appearance of MgS dust is more delayed.

This work builds on the earlier works by adding the rest of
the IRS observations and comparing the samples as a whole. 
Section~2 describes the samples of carbon stars in the 
Magellanic Clouds and the Galaxy and the spectroscopic
and photometric data.  Section~3 describes the data analysis, 
and Section~4 presents the results.  In Section~5, we discuss 
the issues raised, and in Section~6, summarize our findings.  
Appendix~A presents useful relations between infrared filter 
sets and colors for carbon stars.  Appendices~B and C present 
new periods for several sources based on the analysis of 
multi-epoch photometry in the near-IR and mid-IR, 
respectively.

\section{Observations \label{s.obs}} % Sec. 2.0

\subsection{Samples and targets} % Sec. 2.1

%Table 1 - target, alias, ra, dec, ref, period, ref, program

\begin{deluxetable*}{llrrlrlr} % Table 1
\tablecolumns{8}
\tablewidth{0pt}
\tablenum{1}
\tablecaption{The {\it Spitzer} sample of Magellanic carbon stars}
\label{t.sample}
\tablehead{
  \colhead{ } & \colhead{ } & \colhead{RA} & \colhead{Decl.} &
  \colhead{Position} & \colhead{Period} & \colhead{Period} & 
  \colhead{Program} \\
  \colhead{Target} & \colhead{Alias} & \multicolumn{2}{c}{(J2000)} &
  \colhead{reference} & \colhead{(days)} & 
  \colhead{reference\tablenotemark{a}} & \colhead{identifier\tablenotemark{b}}
}
\startdata
GM 780             &                  &   8.905255 & $-$73.165604 & 2MASS &     611 & OGLE    &  3505\\
MSX SMC 091        &                  &   9.236293 & $-$72.421547 & 2MASS & \nodata & \nodata &  3277\\
MSX SMC 062        &                  &  10.670455 & $-$72.951599 & 2MASS &     550 & OGLE    &  3277\\
MSX SMC 054        &                  &  10.774604 & $-$73.361282 & 2MASS &     396 & OGLE    &  3277\\
2MASS J004326      &                  &  10.860405 & $-$73.445366 & 2MASS &     330 & R05     & 50240\\
MSX SMC 044        &                  &  10.914901 & $-$73.249336 & 2MASS &     440 & OGLE    &  3277\\
MSX SMC 105        &                  &  11.258941 & $-$72.873428 & 2MASS &     668 & OGLE    &  3277\\
MSX SMC 036        &                  &  11.474785 & $-$73.394775 & 2MASS &     553 & OGLE    &  3277\\
GB S06             & MSX SMC 060      &  11.668442 & $-$73.279793 & 2MASS &     435 & OGLE    &  3277\\
MSX SMC 200        &                  &  11.711623 & $-$71.794250 & 2MASS &     426 & OGLE    &  3277
\enddata
\tablenotetext{a}{OGLE = \citet[][Optical Gravitational 
Lensing Experiment]{sos09,sos11}; GO7 = \cite{gro07};
G09 = \cite{gro09}; K10 = \cite{kam10}; N00 = \cite{nis00}; 
R05 = \cite{rai05}; S06 = \cite{slo06}; W89 = \cite{whi89}; 
W03 = \cite{whi03}; Z06 = \cite{zij06}; App.~B and C = the
appendices in this paper.}
\tablenotetext{b}{See Table~\ref{t.pid}.}
\tablecomments{Table~\ref{t.sample} is published in its 
entirety in the electronic edition of the Astrophysical 
Journal.  A portion is shown here for guidance regarding its 
form and content.}
\end{deluxetable*}

Table~\ref{t.sample} lists the 144 objects in the LMC and 40 
in the SMC observed with the IRS and identified as carbon 
stars.  Our objective was to build a sample including all of 
the carbon-rich AGB stars observed by the IRS while excluding 
post-AGB objects.  Our selection criteria are based on the 
infrared (IR) spectral classification of \citet[][see 
Section~\ref{s.sws} below]{kra02}.  Generally, spectra are 
considered carbon-rich if they show SiC dust emission at
$\sim$11.3~\mum\ or the acetylene absorption bands at 7.5 
and/or 13.7~\mum.  The 26--30~\mum\ feature attributed to MgS 
dust appears in many of the spectra and helps to confirm 
their carbon-rich nature.

\cite{gm85} identified MgS dust as the carrier of the 
26--30~\mum\ feature, but some doubt has been cast on that 
identification due to abundance constraints 
\citep[e.g.][]{zha09}.  \cite{slo14} examined these and other 
concerns about MgS and concluded that it remains a strong 
candidate for the carrier of the feature.  
Section~\ref{s.layer} investigates this issue, but in the
meantime, we will refer to this feature as the MgS
feature for convenience.

We excluded objects if their spectra peak at wavelengths 
above $\sim$20~\mum\ (in $F_{\nu}$ units), because for 
carbon-rich sources, that indicates that they have
evolved past the AGB.  These excluded objects are part
of the carbon-rich post-AGB sample studied by \cite{slo14}.

Two other classes of post-AGB objects have warmer spectral 
energy distributions (SEDs) peaking shortward of 
$\sim$20~\mum\ and are also excluded.  R~CrB stars have 
smooth spectra showing only emission from amorphous carbon 
dust and no other significant spectral features.  
\cite{kra05} examined two in the SMC, but the IRS sample 
contains additional R CrB stars (Clayton et al., in 
preparation).  The other class of warm post-AGB objects 
includes two sources showing absorption bands from 
aliphatic hydrocarbons\footnote{I.e.\ non-aromatic species 
such as C$_2$H$_6$, longer chains, and other similar 
molecules.}: SMP~LMC~011 \citep{jbs06}, and MSX~SMC~029 
\citep{kra06}.

\begin{deluxetable*}{rll} % Table 2
\tablecolumns{3}
\tablewidth{0pt}
\tablenum{2}
\tablecaption{Spectroscopic {\it Spitzer} programs that 
observed Magellanic carbon stars}
\label{t.pid}
\tablehead{
  \colhead{Program ID} & \colhead{Project leaders} & \colhead{Description}
}
\startdata
  200 & J.~R.~Houck \& G.~C.~Sloan       & Evolved stars in the LMC and SMC\\
 1094 & F.~Kemper                      & AGB evolution in the LMC (and Galaxy)\\
 3277 & M.~P.~Egan \& G.~C.~Sloan        & {\it MSX}-based sample in the SMC\\
 3426 & J.~H.~Kastner \& C.~L.~Buchanan  & Bright infrared sources in the LMC\\
 3505 & P.~R.~Wood \& A.~A.~Zijlstra     & AGB stars in the LMC and SMC\\
 3591 & F.~Kemper                        & AGB evolution in the LMC\\
37088 & R.~Sahai                         & Embedded carbon stars in the LMC\\
40650 & L.~W.~Looney \& R.~A.~Gruendl    & YSOs and red sources in the LMC\\
40519 & A.~G.~G.~M.~Tielens \& F.~Kemper & Extended the IRS sample in the LMC\\
50240 & G.~C.~Sloan \& K.~E.~Kraemer     & Extended the IRS sample in the SMC\\
50338 & M.~Matsuura                      & 
        Carbon-rich post-AGB candidates in the LMC
\enddata
\end{deluxetable*}

A variety of {\it Spitzer} observing programs contributed to 
the present sample of carbon stars (see Table~\ref{t.pid}).
Each program comes with its own selection criteria and 
biases, which means that the present sample is neither 
uniform nor complete.  However, thanks to the efforts of the 
designers of all the previous programs, the present sample 
probes carbon stars spanning a wide range of luminosities, 
colors, and evolutionary stages.

The first two programs listed in Table~\ref{t.pid} were
developed prior to launch as part of the guaranteed-time 
observations (GTOs).  Both sampled a variety of evolved 
stars, with Program 200 focused on supergiants and AGB stars 
identified as variable stars in the optical and near-IR in 
both the SMC and LMC.  Program 1094 covered a range of 
post-main-sequence evolutionary stages in the LMC (and the
Galaxy).  

The next four programs were Guest Observer (GO) programs
in Cycle 2 of the mission.  Program 3277 started with the 
mid-IR sources in the SMC identified by the {\it Mid-course 
Space Experiment} \citep[{\it MSX};][]{pri01} to generate 
a sample spanning color-color space in the near- and mid-IR.  
Program 3426 sampled a randomly chosen subset of the 
brightest 150 sources in the LMC.  Program 3505 identified 
AGB stars in several previous infrared surveys with the help 
of near-IR color-magnitude diagrams (CMDs).  It covered both 
the LMC and SMC.  Program 3591 extended Program 1094 to a 
larger sample in the LMC.

More programs followed later in the {\it Spitzer} mission.
Program 37088 concentrated on heavily embedded and thus more 
evolved stars in the LMC.  Program 40650 targeted about 300 
candidate young stellar objects (YSOs) in the LMC, as well as 
several sources described as extremely red objects (EROs) and 
subsequently identified as deeply embedded carbon stars 
\citep{gru08}.  The latter group is included in our sample.  
Program 40519 expanded on all previous programs studying the 
LMC by observing about 200 targets covering undersampled 
regions of color-color and color-magnitude space.  It 
included evolved stars and YSOs, relied on the mid-IR 
photometry from the SAGE survey of the LMC \citep[Surveying 
the Agents of Galactic Evolution;][]{mei06}, and is known as 
SAGE-Spec.  This program added a number of faint carbon-rich 
targets to the sample.  Program 50240 expanded the targets 
observed in the SMC in a similar manner.  Finally, Program 
50338 concentrated on post-AGB candidates in the LMC.

\subsection{Adopted distances} % Sec. 2.2

The known distances to the LMC and SMC enable us to study
the total luminosities of the stars in our sample.  We
adopt distance moduli for the LMC and SMC of 18.5 and 18.9, 
respectively.  Measurements of the distance modulus to the 
LMC cluster around our adopted value.  The review by 
\cite{fea13} leads to 18.50; another good example of a recent 
result is 18.49 $\pm$ 0.05 \citep{pie13}.  Similarly, our 
adopted distance modulus to the SMC is close to the mean 
value reported by \cite{rub15}, 18.91 $\pm$ 0.02.  The 
distances to these galaxies are only important to our 
determinations of absolute bolometric magnitudes, and because 
these have uncertainties larger than 0.1 magnitudes, we are 
not concerned with precision in the distance modulus to 
further significant figures.

The SMC has a complicated structure, with considerable depth 
along the line of sight \citep{gar91}.  Most of the stars 
reside in the Bar region, where the mean distance modulus to 
different regions can vary from $\sim$18.85 to 19.02 
\citep{rub15}.  Typical depths in this region are only 
2--3~kpc ($\Delta$($m$$-$$M$)$\la$ 0.1), but in the extended 
Wing region to the east, \cite{nid13} report depths up to 
$\sim$20~kpc ($\Delta$($m$$-$$M$)$\la$ 0.8).  \cite{rub15} 
find mean distance moduli in the eastern regions between 
$\sim$18.65 and 18.8.  Most of the targets are in the Bar, 
and we will assume they are at our canonical distance
modulus (18.9).

We have several targets in the cluster NGC~419, which is in
the direction of the SMC.  \cite{gla08} give a distance 
modulus of 18.50, the same as for the LMC and about 10 kpc in 
front of the SMC.  They describe how the model which best 
fits the optical CMD gives ($m-M$)$_o$ = 18.75, while other 
models give values of 18.60 and 18.94.  It is unclear what 
justifies the reported value of 18.50, and as discussed 
below, a value of 18.90 is more consistent with the IR data 
(see Section~\ref{s.mbol}).

\subsection{IRS observations} % Sec. 2.3

The IRS observed the majority of the sources in the standard
low-resolution nod sequence, most of them with both the
Short-Low (SL) and Long-Low (LL) modules.  SL and LL provide
spectral coverage from 5--14~\mum\ and 14--37~\mum,
respectively, with a resolution ($\lambda$/$\Delta\lambda$)
of $\sim$80--120.  Some of the fainter sources were observed 
only in SL.  Generally, sources were observed in each of the
two SL apertures, providing first- and second-order spectra,
then the two LL apertures.  Each source was observed in two
nod positions in each aperture.  Thus a full low-resolution
spectrum requires eight pointings.

The carbon-rich EROs observed by \cite{gru08} used the SL 
module, along with Short-High (SH) and Long-High (LH), which 
cover 10--19 and 19--36~\mum, respectively, at a resolution 
of $\sim$300.

The reduction and calibration of the low-resolution spectra 
followed the standard Cornell algorithm \citep[see][for a
more detailed description]{slo15a}. The Cornell 
processing begins with the flatfielded images provided by 
version S18.18 of the pipeline at the {\it Spitzer} Science 
Center (SSC).  Bad pixels in each image are identified and 
replaced with values based on their neighbors.  All images 
for a given nod position are averaged, then a suitable 
background image is subtracted from the target image.  In SL, 
the default is an aperture differerence, where the background 
for a given target is the image with the target in the other 
aperture, while in LL, the default is a nod difference, with 
the target in the same aperture, but the other nod.  In some 
cases, the design of the observation or structure or other 
sources in the background forced a change from the default.

We deviated from the reduction sequence described by
\cite{slo15a} by using an optimal-extraction algorithm to 
extract spectra from the images \citep{leb10}.  This method 
fits a supersampled point-spread function (PSF) to the 
spatial profile in each row of the spectral image and can 
improve the SNR up to 80\%.  It proved 
essential to generating usable spectra for the faintest 
sources in our sample.  However, the method only works for 
point sources.  For extended or mispointed objects, we used 
the default extraction algorithm which simply sums data in 
each wavelength element with no weighting.

Our pipeline continues with the standard step of combining 
spectra from the two nods and comparing the data to identify 
and reject spikes or divots which had survived the cleaning 
process.  Spectral segments are then stitched together, 
shifting spectra upwards multiplicatively to the presumably 
best-centered segment.  Finally, extraneous data are removed 
from the ends of each segment.

Several previous papers on Magellanic carbon stars used the
above algorithm, but without the optimal-extraction method 
which came later \citep{slo06,zij06,lag07,lei08,slo08,kem10}.
The optimal extraction has generally improved the quality of 
the spectra.

\subsection{Photometric data \label{s.photo}} % Sec. 2.4

\begin{deluxetable*}{lllllllr} % Table 3
\tablecolumns{8}
\tablewidth{0pt}
\tablenum{3}
\tablecaption{Optical and near-infrared photometry}
\label{t.optnir}
\tablehead{
  \colhead{ } & \colhead{$U$} & \colhead{$B$} & \colhead{$V$} &
  \colhead{$I$} & \colhead{$J$} & \colhead{$H$} & \colhead{$K$} \\
  \colhead{Target} & \colhead{(mag.)} & \colhead{(mag.)} & \colhead{(mag.)} &
  \colhead{(mag.)} & \colhead{(mag.)} & \colhead{(mag.)} & \colhead{(mag.)} 
}
\startdata
GM 780             & \phm{22.222}   \nodata  & \phm{22.222}   \nodata  &   16.760 $\pm$ \nodata  &   14.235 $\pm$    0.040 &   12.098 $\pm$    0.740 &   10.658 $\pm$    0.564 &    9.934 $\pm$    0.402\\
MSX SMC 091        & \phm{22.222}   \nodata  & \phm{22.222}   \nodata  & \phm{22.222}   \nodata  &   15.781 $\pm$    0.060 &   13.764 $\pm$    0.374 &   12.259 $\pm$    0.284 &   10.829 $\pm$    0.191\\
MSX SMC 062        & \phm{22.222}   \nodata  & \phm{22.222}   \nodata  & \phm{22.222}   \nodata  &   16.985 $\pm$ \nodata  &   13.261 $\pm$    1.398 &   11.418 $\pm$    0.159 &   10.281 $\pm$    0.769\\
MSX SMC 054        & \phm{22.222}   \nodata  & \phm{22.222}   \nodata  &   21.430 $\pm$ \nodata  &   20.383 $\pm$ \nodata  &   16.536 $\pm$    0.854 &   14.449 $\pm$    0.182 &   12.281 $\pm$    0.369\\
2MASS J004326      & \phm{22.222}   \nodata  & \phm{22.222}   \nodata  &   18.658 $\pm$ \nodata  &   15.170 $\pm$    0.040 &   13.041 $\pm$    0.117 &   11.857 $\pm$    0.108 &   10.970 $\pm$    0.151\\
MSX SMC 044        &   19.762 $\pm$    0.135 &   19.230 $\pm$    0.039 &   19.920 $\pm$    0.045 &   16.406 $\pm$    1.882 &   13.154 $\pm$    1.336 &   12.178 $\pm$    1.822 &   10.492 $\pm$    0.943\\
MSX SMC 105        & \phm{22.222}   \nodata  &   22.447 $\pm$    0.391 &   21.099 $\pm$    0.303 &   18.370 $\pm$    0.081 &   15.151 $\pm$    0.446 &   13.082 $\pm$    0.382 &   11.245 $\pm$    0.290\\
MSX SMC 036        & \phm{22.222}   \nodata  & \phm{22.222}   \nodata  & \phm{22.222}   \nodata  &   19.462 $\pm$    0.090 &   15.170 $\pm$    1.013 &   13.553 $\pm$    0.898 &   11.743 $\pm$    0.632\\
GB S06             & \phm{22.222}   \nodata  & \phm{22.222}   \nodata  &   20.137 $\pm$ \nodata  &   18.114 $\pm$ \nodata  &   15.348 $\pm$    0.364 &   13.146 $\pm$    0.283 &   11.211 $\pm$    0.242\\
MSX SMC 200        & \phm{22.222}   \nodata  & \phm{22.222}   \nodata  &   21.203 $\pm$ \nodata  &   17.557 $\pm$    0.200 &   14.612 $\pm$    0.899 &   12.724 $\pm$    0.804 &   11.371 $\pm$    0.413
\enddata
\tablecomments{Table~\ref{t.optnir} is published in its 
entirety in the electronic edition of the Astrophysical 
Journal.  A portion is shown here for guidance regarding its 
form and content.}
\end{deluxetable*}

\begin{deluxetable*}{lrrrrrr} % Table 4
\tablecolumns{7}
\tablewidth{0pt}
\tablenum{4}
\tablecaption{Mid-infrared photometry and bolometric magnitudes}
\label{t.mir}
\tablehead{
  \colhead{ } & \colhead{[3.6]} & \colhead{[4.5]} & \colhead{[5.8]} &
  \colhead{[8.0]} & \colhead{[24]} & \colhead{$M_{{\rm bol}}$} \\
  \colhead{Target} & \colhead{(mag.)} & \colhead{(mag.)} &
  \colhead{(mag.)} & \colhead{(mag.)} & \colhead{(mag.)} & \colhead{(mag.)}
}
\startdata
GM 780             &    8.738 $\pm$    0.454 &    8.288 $\pm$    0.413 &    8.283 $\pm$    0.064 &    7.916 $\pm$    0.066 &    6.771 $\pm$    0.368 & $-$5.90\\
MSX SMC 091        &    9.602 $\pm$    0.366 &    8.976 $\pm$    0.322 &    8.260 $\pm$    0.008 &    7.864 $\pm$    0.029 &    7.328 $\pm$    0.060 & $-$4.88\\
MSX SMC 062        &    8.941 $\pm$    0.266 &    8.375 $\pm$    0.304 &    7.639 $\pm$    0.026 &    7.092 $\pm$    0.023 &    6.272 $\pm$    0.178 & $-$5.50\\
MSX SMC 054        &    9.940 $\pm$    0.327 &    9.037 $\pm$    0.313 &    8.408 $\pm$    0.201 &    7.771 $\pm$    0.142 &    6.595 $\pm$    0.142 & $-$4.58\\
2MASS J004326      &   10.495 $\pm$    0.400 &   10.476 $\pm$    0.257 &   10.196 $\pm$    0.095 &    9.615 $\pm$    0.061 &    9.316 $\pm$    0.056 & $-$4.56\\
MSX SMC 044        &    9.300 $\pm$    0.256 &    8.661 $\pm$    0.230 &    8.189 $\pm$    0.191 &    7.713 $\pm$    0.155 &    7.216 $\pm$    0.077 & $-$5.29\\
MSX SMC 105        &    9.312 $\pm$    0.129 &    8.544 $\pm$    0.151 &    8.007 $\pm$    0.079 &    7.370 $\pm$    0.078 &    5.928 $\pm$    0.220 & $-$5.04\\
MSX SMC 036        &    9.698 $\pm$    0.314 &    8.790 $\pm$    0.274 &    8.051 $\pm$    0.199 &    7.515 $\pm$    0.160 &    6.529 $\pm$    0.131 & $-$4.62\\
GB S06             &    8.666 $\pm$    0.259 &    7.927 $\pm$    0.373 &    6.922 $\pm$    0.157 &    6.241 $\pm$    0.182 &    4.874 $\pm$    0.098 & $-$5.62\\
MSX SMC 200        &    9.298 $\pm$    0.180 &    8.806 $\pm$    0.198 &    8.362 $\pm$    0.062 &    7.894 $\pm$    0.030 &    7.276 $\pm$    0.206 & $-$4.81
\enddata
\tablecomments{Table~\ref{t.mir} is published in its entirety 
in the electronic edition of the Astrophysical Journal.  A 
portion is shown here for guidance regarding its form and 
content.}
\end{deluxetable*}

For all of our targets, we have constructed SEDs based on 
multi-epoch photometry in the optical, near-IR, and mid-IR.  

The mid-IR data come from the SAGE survey of the LMC 
\citep{mei06} and the SAGE-SMC survey for the SMC 
\citep{gor11}, which incorporates the S$^3$MC survey 
\citep[{\it Spitzer} Survey of the SMC;][]{bol07}.  The 
SAGE surveys give two epochs, spaced approximately three 
months apart, for the Infrared Array Camera (IRAC) and the 
Multi-Imaging Photometer for {\it Spitzer} (MIPS).  The 
S$^3$MC survey provides one additional epoch in the core of 
the SMC.  We use all four IRAC filters at 3.6, 4.5, 5.8, and
8.0~\mum\ and the MIPS 24~\mum\ filter.  The SAGE-VAR survey 
adds four epochs from the Warm {\it Spitzer} Mission at 3.6 
and 4.5~\mum\ for portions of the LMC and SMC \citep{rie15}.  
We found matches with the SAGE-VAR survey for 61 of our 
carbon stars in the LMC and 26 in the SMC.

We also used additional epochs at 3.4 and 4.6~\mum\ from the 
{\it Wide-field Infrared Survey Experiment} 
\citep[{\it WISE};][]{wri10} and the {\it NEOWISE} reactivation 
mission \citep{mai14}.\footnote{We used the AllWISE 
Multiepoch Photometry Table and the NEOWISE-R Single Exposure 
Source Table available online at the Infrared Science Archive 
(IRSA).}  For all {\it WISE} data, we utilized the 
multi-epoch data tables and collapsed them to one epoch 
approximately every six months.  The publicly available data 
typically give us four epochs, two in 2010 and two in 2014.  

Applying the color corrections derived for carbon stars in 
Appendix~A allows us to convert W1 (3.4~\mum) to [3.6] and 
W2 (4.6~\mum) to [4.5].  The combination of the SAGE surveys,
including SAGE-VAR, and the {\it WISE} data can give us 10 or 
more epochs at 3.6 and 4.5~\mum\ for some of the stars in our
sample.

The {\it WISE} survey is less sensitive and has lower angular
resolution than the SAGE data, forcing us to reject some
questionable data.  We rejected both W1 and W2 for NGC~419
LE~35 and MIR~1 due to crowding.  For IRAS~05026 and 
IRAS~05042, the multi-epoch data indicated a detection at
W1, but the {\it WISE} images of the field showed nothing, even
though the sources could be seen in W2.  The W1$-$W2 colors
were inconsistent with the SAGE data, so we rejected the W1
data but kept W2.
% rejections - cf notes 3 Nov 15 p 2

Near-IR photometry comes from the 2MASS survey, and the 
deeper 2MASS-6X survey provides a second epoch at $J$, $H$, 
and $K_s$ \citep{skr06, cut06}.  Additional epochs come from 
the Deep Near-IR Survey of the Southern Sky (DENIS) at 
$J$ and $K_s$ \citep{cio00} and the IR Survey Facility (IRSF) 
at $J$, $H$, and $K_s$ \citep{kat07}.  

In the optical, we relied on the Magellanic Clouds 
Photometric Survey (MCPS) at $U$, $B$, $V$, and $I$ 
\citep{zar02, zar04}.  DENIS adds data at $I$.  Additional 
mean magnitudes at $V$ and $I$ in the LMC come from the 
OGLE-III Shallow Survey \citep{ula13}.  Where possible, we 
replaced the $V$ and $I$ data with mean magnitudes from the 
OGLE-III surveys of the Magellanic Clouds, which also give 
pulsation periods and amplitudes \citep{sos09,sos11}.

We searched for matches to the IRS sources first with the 
SAGE surveys, using positions estimated from the pointing 
of the IRS.  The optimal extraction algorithm estimates 
the position of a source along a given slit (i.e., in the 
cross-dispersion direction).  For most observations, the 
nearly perpendicular orientation of the SL and LL slits helps 
to constrain the position of the source, despite the widths 
of the slits ($\sim$3$\farcs$6 and 10$\farcs$6, 
respectively).  Where possible, we used matches to IRAC to 
update the initial IRS-based coordinates, then searched for 
2MASS counterparts, again updating positions where possible, 
and then searched the other catalogs.  The positions in 
Table~\ref{t.sample} are the result.

The median offset from the estimated IRS position to IRAC
is 0$\farcs$37, consistent with a typical IRS pointing
error of 0$\farcs$4.  The median offset between the IRAC and
2MASS positions is 0$\farcs$20.  Comparing positions in the
IRAC and MIPS catalogs, we find a median offset of 
0$\farcs$20.
% pointing info - cf notes 15 Sep 15 and 22 Sep 15

For the most embedded sources, the expected photometry in
the near-IR and at 3.6~\mum\ is near or below the detection 
limit, making it possible to mismatch the photometry to
blue sources close to the expected positions of our carbon
stars.  We examined the environments of the most embedded 
sources, searching for targets within 1$\arcmin$ to estimate 
the source density.  For IRAC, the source densities are 
1.1--4.8$\times$10$^{-3}$, and for 2MASS-6X, they are higher, 
7.0--8.2 $\times$10$^{-3}$.  Our maximum search radius in the 
LMC was 1$\farcs$25, which gives a search field of 4.9 square 
arcsec and makes the odds of a mismatch for a given source 
0.5--2.4\%.  For 2MASS-6X, the odds of a mismtach are 
3.4--4.0\% for a given source.  Thus, mismatches are 
unlikely.

However, the above calculations were for field stars.  For 
stars in clusters, we compared our near-IR results to the 
photometry reported by \cite{vl05}, and we used these data to 
replace the $J$ magnitude of one source, NGC~1978 MIR~1.

Tables~\ref{t.optnir} and \ref{t.mir} present the resulting
photometry.  For each photometric filter, the magnitude is a
mean magnitude, and the uncertainty is the standard
deviation of the data.  When only one epoch of data in a 
given filter was available, we left the reported uncertainty
undefined.  While we can have over ten epochs at 3.6 and 
4.5~\mum, at longer wavelengths we are limited to three 
epochs in the heart of the SMC and two epochs in the LMC and 
the outskirts of the SMC.

\subsection{Galactic comparison spectra \label{s.sws}} % Sec. 2.5

We also consider a Galactic control sample using spectra from
the Short-Wavelength Spectrometer \citep[SWS;][]{deg96} on the
{\it Infrared Space Observatory} \citep[{\iso};][]{kes96}.
\cite{lei08} presented the original list of 34 carbon stars
used in earlier comparisons to Magellanic samples.  The
present sample includes eight additional objects, for a total 
of 42.

The comparison SWS sample was chosen from spectra classified 
as carbon stars by \cite{kra02}.  Their classification scheme 
assigns spectra to groups based on the overall shape of the 
spectrum, with naked stars in Group 1, stellar spectra 
showing some dust emission in Group 2, spectra dominated by 
warm dust in Group 3, and spectra dominated by cold dust (so 
that the spectrum peaks past $\sim$20~\mum) in Groups 4 and 
5.  Spectra are assigned to subgroups based on the dominant 
features in the spectra.  The carbon stars are those with the 
following classifications:  1.NC (naked star with carbon-rich 
molecular absorption bands), 2.CE and 3.CE (optically thin 
carbon-rich dust emission), and 3.CR (reddened spectra from 
optically thick carbon-rich dust emission).  The original 
sample of 34 described by \cite{lei08} did not include any 
naked carbon stars; we have added three 1.NC sources.  Nor 
did it consider the 23 slower SWS scans, which adds three 
more sources.  The additional two sources were simply 
overlooked before.

\section{Analysis \label{s.ana}} % Sec. 3.0

\subsection{The Manchester method \label{s.mm}} % Sec. 3.1

\begin{figure} % Fig. 1
\includegraphics[width=3.4in]{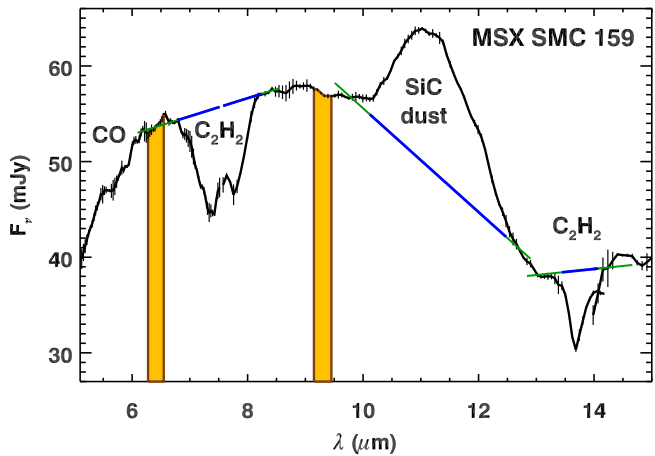} % figmm1.eps
\caption{The Manchester method applied to the 5--15~\mum\
spectrum of MSX~SMC~159.  The [6.4]$-$[9.3] color provides
an estimate of the relative contributions of stellar 
photosphere and amorphous carbon dust in two spectral regions
relatively free of absorption or emission features.  Line 
segments are used to estimate the continuum under or over 
the C$_2$H$_2$ absorption band at 7.5~\mum, the Q branch
of the C$_2$H$_2$ band at 13.7~\mum, and the SiC dust
emission feature at $\sim$11.3~\mum.\label{f.mm1}}
\end{figure}

\begin{figure} % Fig. 2
\includegraphics[width=3.4in]{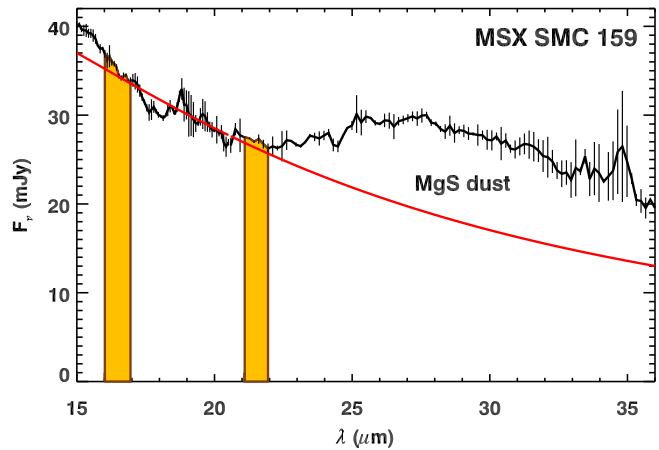} % figmm2.eps
\caption{The Manchester method applied to the 15--37~\mum\
spectrum of MSX~SMC~159.  The [16.5]$-$[21.5] color estimates
the continuum temperature at longer wavelengths in order to
estimate the continuum under the MgS dust emission feature at 
longer wavelengths.\label{f.mm2}}
\end{figure}

\begin{deluxetable*}{lrrrrrc} % Table 5
\tablecolumns{7}
\tablewidth{0pt}
\tablenum{5}
\tablecaption{Spectroscopic data --- IRS sample}
\label{t.irsdat}
\tablehead{
  \colhead{ } & \colhead{Eq.\ width of} & \colhead{$F_{SiC}$~/} & 
  \colhead{[6.4]$-$[9.3]} & \colhead{[16.5]$-$[21.5]} & \colhead{$F_{MgS}$~/} &
  \colhead{IR spectral}  \\
  \colhead{Target} & \colhead{C$_2$H$_2$ at 7.5~um} & \colhead{continuum} & 
  \colhead{(mag.)} & \colhead{(mag.)} & \colhead{continuum} &
  \colhead{classification} 
}
\startdata
GM 780             &    0.162 $\pm$    0.010 &    0.316 $\pm$    0.006 &    0.403 $\pm$    0.008 &    0.165 $\pm$    0.024 & \nodata \phm{0.222}     & CE2\\
MSX SMC 091        &    0.146 $\pm$    0.015 &    0.147 $\pm$    0.007 &    0.528 $\pm$    0.007 &    0.073 $\pm$    0.025 & \nodata \phm{0.222}     & CE2\\
MSX SMC 062        &    0.117 $\pm$    0.006 &    0.062 $\pm$    0.003 &    0.675 $\pm$    0.008 &    0.174 $\pm$    0.024 & \nodata \phm{0.222}     & CE3\\
MSX SMC 054        &    0.177 $\pm$    0.006 &    0.179 $\pm$    0.003 &    0.756 $\pm$    0.004 &    0.101 $\pm$    0.016 &    0.191 $\pm$    0.021 & CE3\\
2MASS J004326      &    0.196 $\pm$    0.007 &    0.080 $\pm$    0.006 &    0.234 $\pm$    0.019 & \nodata \phm{0.222}     & \nodata \phm{0.222}     & CE1\\
MSX SMC 044        &    0.050 $\pm$    0.005 &    0.024 $\pm$    0.004 &    0.517 $\pm$    0.006 & $-$0.101 $\pm$    0.046 & \nodata \phm{0.222}     & CE2\\
MSX SMC 105        &    0.184 $\pm$    0.008 &    0.157 $\pm$    0.005 &    0.866 $\pm$    0.003 &    0.226 $\pm$    0.022 &    0.254 $\pm$    0.029 & CE3\\
MSX SMC 036        &    0.234 $\pm$    0.006 &    0.088 $\pm$    0.004 &    0.774 $\pm$    0.005 &    0.286 $\pm$    0.017 &    0.279 $\pm$    0.021 & CE3\\
GB S06             &    0.113 $\pm$    0.005 &    0.058 $\pm$    0.002 &    0.969 $\pm$    0.003 &    0.284 $\pm$    0.020 &    0.358 $\pm$    0.025 & CE4\\
MSX SMC 200        &    0.032 $\pm$    0.007 &    0.010 $\pm$    0.003 &    0.443 $\pm$    0.008 &    0.067 $\pm$    0.029 & \nodata \phm{0.222}     & CE2
\enddata
\tablecomments{Table~\ref{t.irsdat} is published in its 
entirety in the electronic edition of the Astrophysical 
Journal.  A portion is shown here for guidance regarding its 
form and content.}
\end{deluxetable*}

\begin{deluxetable*}{llrrrrrc} % Table 6
\tablecolumns{8}
\tablewidth{0pt}
\tablenum{6}
\tablecaption{Spectroscopic data --- SWS sample}
\label{t.swsdat}
\tablehead{
  \colhead{ } & \colhead{ } & \colhead{Eq.\ width of} & \colhead{$F_{SiC}$~/} &
  \colhead{[6.4]$-$[9.3]} & \colhead{[16.5]$-$[21.5]} & \colhead{$F_{MgS}$~/} & \colhead{IR spectral} \\
  \colhead{Target} & \colhead{Alias} & \colhead{C$_2$H$_2$ at 7.5~um} & \colhead{continuum} &
  \colhead{(mag.)} & \colhead{(mag.)} & \colhead{continuum} & \colhead{classification}
}
\startdata
WZ Cas            & \nodata   &    0.347 $\pm$ 0.007 & $-$0.022 $\pm$ 0.002 &    0.029 $\pm$ 0.002 &    0.438 $\pm$ 0.004 & \nodata \phm{0.222}     & CE0\\
VX And            & \nodata   &    0.238 $\pm$ 0.001 &    0.182 $\pm$ 0.001 &    0.123 $\pm$ 0.003 &    0.067 $\pm$ 0.010 & \nodata \phm{0.222}     & CE1\\
HV Cas            & \nodata   &    0.090 $\pm$ 0.001 &    0.226 $\pm$ 0.001 &    0.353 $\pm$ 0.002 & $-$0.139 $\pm$ 0.004 & \nodata \phm{0.222}     & CE2\\
R Scl             & \nodata   &    0.290 $\pm$ 0.002 &    0.205 $\pm$ 0.001 &    0.273 $\pm$ 0.002 &    0.266 $\pm$ 0.002 &    0.292 $\pm$    0.004 & CE1\\
R For             & \nodata   &    0.125 $\pm$ 0.001 &    0.279 $\pm$ 0.001 &    0.361 $\pm$ 0.001 &    0.131 $\pm$ 0.003 & \nodata \phm{0.222}     & CE2\\
AFGL 341          & V596 Per  &    0.063 $\pm$ 0.001 &    0.073 $\pm$ 0.001 &    1.665 $\pm$ 0.005 &    0.407 $\pm$ 0.008 &    0.411 $\pm$    0.010 & CE5\\
CIT 5             & V384 Per  &    0.065 $\pm$ 0.001 &    0.305 $\pm$ 0.002 &    0.682 $\pm$ 0.001 &    0.323 $\pm$ 0.002 &    0.186 $\pm$    0.002 & CE3\\
U Cam             & \nodata   &    0.078 $\pm$ 0.001 &    0.327 $\pm$ 0.001 &    0.107 $\pm$ 0.002 &    0.157 $\pm$ 0.004 & \nodata \phm{0.222}     & CE1\\
W Ori             & \nodata   &    0.064 $\pm$ 0.001 &    0.229 $\pm$ 0.001 &    0.055 $\pm$ 0.001 & $-$0.054 $\pm$ 0.004 &    0.137 $\pm$    0.002 & CE1\\
IRC $-$10095      & V1187 Ori &    0.277 $\pm$ 0.002 &    0.192 $\pm$ 0.001 &    0.163 $\pm$ 0.003 &    0.196 $\pm$ 0.008 &    0.784 $\pm$    0.012 & CE1\\
\enddata
\tablecomments{Table~\ref{t.swsdat} is published in its
entirety in the electronic edition of the Astrophysical
Journal.  A portion is shown here for guidance regarding its
form and content.}
\end{deluxetable*}

\cite{slo06} and \cite{zij06} introduced the Manchester 
method to extract and compare information uniformly from 
large samples of infrared spectra from carbon stars.  The 
key metric is the [6.4]$-$[9.3] color, which samples the
spectrum at two wavelengths that are relatively free of 
molecular absorption bands and solid-state emission 
features.  This color reddens as the amount of amorphous 
carbon grows above the stellar photosphere.  \cite{gro07}
found that the [6.4]$-$[9.3] color increased linearly with 
the log of the dust-production rate (DPR).  The current 
sample includes bluer sources which deviate from a linear 
relationship, and it requires a more complex formulation 
\citep[e.g.][]{mat09,gul12,rie12}.  All of these formulations 
are based on a tight relationship between color and DPR and 
reinforce our assumption that the [6.4]$-$[9.3] color is a 
good proxy for DPR.

The Manchester method measures the equivalent width of the 
acetylene absorption bands at 7.5 and 13.7~\mum\ by fitting 
line segments to the continua to either side.  The 13.7~\mum\ 
band is actually just the Q branch of a broader feature, and 
we will focus on the stronger 7.5~\mum\ band.  The strength 
of the SiC dust emission feature at $\sim$11.5~\mum\ is 
measured similarly and is reported as a ratio of integrated 
flux in the feature to the integrated flux in the underlying 
continuum.  

We also measure the strength of the 26--30~\mum\ feature
attributed to MgS, but because the IRS spectral coverage
cuts off the long-wavelength side of the feature, we use
a two-step process.  First we estimate a continuum based
on the apparent color of the spectrum at 16.5 and 21.5~\mum.
Then we integrate the spectrum above this estimated
continuum.  If the spectra do not appear to be turning
down at the long-wavelength cut-off, we assume that the
26--30~\mum\ feature is not present or is contaminated in
some way and do not report a value.  As with the SiC
feature, we report the integrated flux in the feature,
divided by the integrated continuum.

Figures~\ref{f.mm1} and \ref{f.mm2} illustrate the 
application of the Manchester method to the spectrum of one 
Magellanic carbon star, MSX~SMC~159.  \cite{slo06} presented 
similar figures for the same star (see their Figures 4 and 
5).  The difference is that we used optimal extraction 
for the spectrum here, with a noticeable improvement in SNR.  

Table~\ref{t.irsdat} presents the results for the IRS sample
of carbon stars in the Magellanic Clouds.  
Table~\ref{t.swsdat} presents the 42 Galactic carbon stars 
observed with the SWS.  The results in Table~\ref{t.swsdat} 
differ from those presented by \cite{lei08} because of 
differences in how the continuum was estimated.  As explained 
in Appendix~A, 
the [6.4]$-$[9.3] colors of two sources, IRAS~04589 and 
IRAS~05306, were artificially reddened because the spectra 
were badly mispointed.  For these, we replaced the 
[6.4]$-$[9.3] colors with values estimated from their 
photometric [5.8]$-$[8] color.

\subsection{Spectral classification \label{s.spclass}} % Sec. 3.2

\begin{figure} % Fig. 3
\includegraphics[width=3.4in]{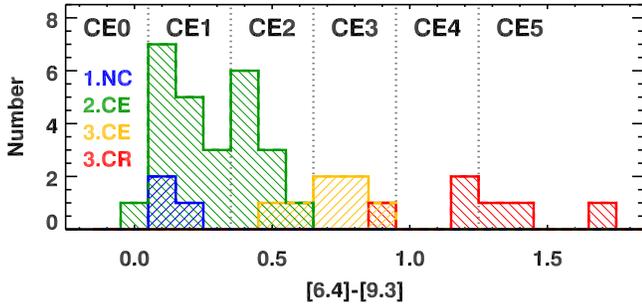} % galhist.eps
\caption{The distribution of the Galactic carbon stars in
our control sample with [6.4]$-$[9.3] color, segregated by
their infrared spectral classifications from \cite{kra02}.
The [6.4]$-$[9.3] color increases nearly monotonically along
the sequence 1.NC--2.CE--3.CE--3.CR.\label{f.galhist}}
\end{figure}

Table~\ref{t.irsdat} also includes infrared spectral 
classifications, which are based on modifications to the 
two-level scheme applied by \cite{kra02} to the SWS data 
(see Section~\ref{s.sws}).  The original scheme envisioned a 
third level of classification for some of the more populated 
subgroups, but this level was adopted only for the silicate 
emission (SE) sources by using the SE indices defined by 
\cite{sp95}.  

The scheme as modified here changes how carbon stars are 
classified.  Previously, they fell into four groups in
the sequence 1.NC--2.CE--3.CE--3.CR as the carbon-rich 
dust shell grew progressively thicker and redder.  The 
new scheme replaces this sequence with CE0--CE5, depending
solely on the [6.4]$-$[9.3] color, with the breaks at a 
color of 0.05 and at intervals of 0.30 from there to 1.25.  

Figure~3 uses the classifications of Galactic carbon stars
by \cite{kra02} to show that the old classifications closely 
follow the [6.4]$-$[9.3] color.  The new sequence shifts the 
boundaries between the subclasses, but presents them in a 
more easily recognizable sequence.

Some of the redder CE5 sources show negative values for the
SiC-to-continuum ratio, because the SiC feature is in 
absorption instead of emission.  For those sources with a
3$\sigma$ detection or better, we have designated them as 
``CA'' instead of ``CE'', for absorption instead of emission.
This classification is analogous to ``SA'' versus ``SE''
introduced for silicates by \cite{sp95}.

\section{Results \label{s.results}} % Sec. 4.0

The present sample has added over 70 sources to the 
previously published samples of Magellanic carbon stars.
The first question to address is whether or not the
larger sample contradicts any earlier conclusions.  We
therefore base our analysis, as before, on comparisons
using the [6.4]$-$[9.3] color as a proxy for DPR.  This
approach compares the properties of the stars and their 
spectra from populations with different metallicities but
similar DPRs.

\subsection{The [6.4]$-$[9.3] color} % Sec. 4.1

\begin{figure} % Fig. 4
\includegraphics[width=3.4in]{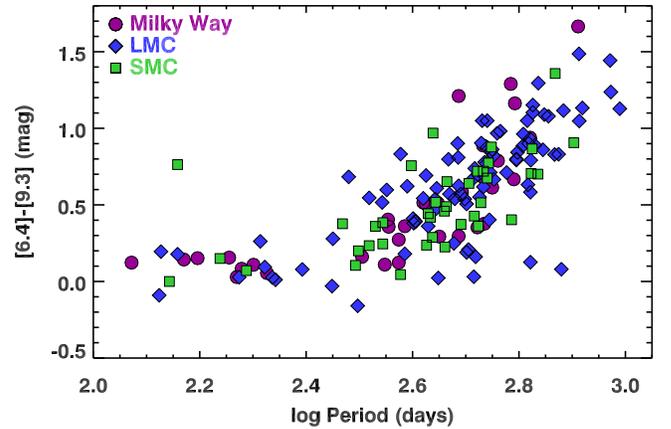} % p_c69_1603.eps
\caption{The [6.4]$-$[9.3] color of the carbon stars in the
Galaxy, LMC, and SMC as a function of pulsation period.  For
periods $\ga$250 days, the colors generally grow redder with 
longer periods, but with considerable scatter.  No dependency 
on metallicity is apparent.\label{f.pc69}}
\end{figure}

Figure~\ref{f.pc69} shows how the [6.4]$-$[9.3] color of the
carbon stars depends on pulsation period.  The larger sample
considered here reinforces the previous conclusions.
First, the dust-production rate, as measured by the 
[6.4]$-$[9.3] color, generally increases with longer 
pulsation periods, once the period exceeds $\sim$250~d.  
Second, the scatter is significant, with a range of DPRs 
possible at a given pulsation period.  Third, the three 
galaxies considered show no obvious differences, indicating 
that metallicity does not have a noticeable effect.

\subsection{Infrared spectral features} % Sec. 4.2

\subsubsection{Silicon carbide dust} % Sec. 4.2.1

\begin{figure} % Fig. 5
\includegraphics[width=3.4in]{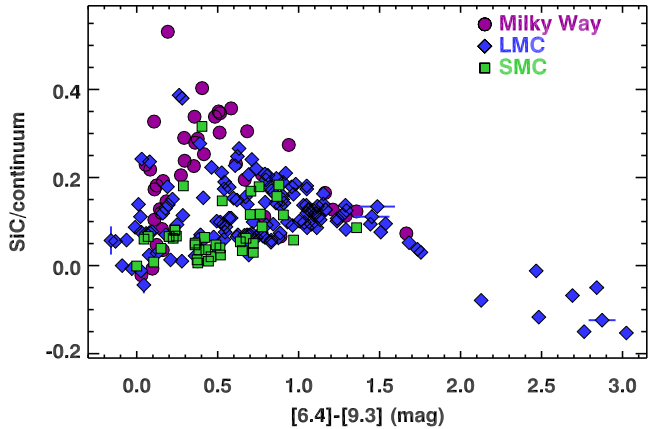} % c69sic_1511.eps
\caption{The strength of SiC dust emission as a function of 
[6.4]$-$[9.3] color.  As the [6.4]$-$[9.3] color increases,
stars follow one of two tracks.  The SiC strength increases
quickly in more metal-rich stars, then turns over and 
decreases.  In metal-poor stars, the SiC strength increases 
more gradually.  Most of the stars with [6.4]$-$[9.3] $\ga$ 
1.5 have negative SiC strengths, indicating absorption.
\label{f.csic}}
\end{figure}

Figure~\ref{f.csic} shows that the dependence of the strength 
of the SiC feature on the [6.4]$-$[9.3] color changes with 
host galaxy.  For the Galactic sources, the SiC dust emission
increases quickly from [6.4]$-$[9.3] = 0 to $\sim$0.4, then 
drops steadily to a color of $\sim$1.7, which is the maximum 
in the Galactic sample.  Most of the carbon stars in the SMC 
sample behave differently, with SiC strength rising more 
gradually with color to [6.4]$-$[9.3] $\sim$ 1.4 for the 
reddest carbon star observed with the IRS in the SMC.  The 
LMC sample is more evenly split between the upper and lower 
sequences.

\cite{slo06} first noticed this metallicity-dependent 
difference when they compared samples from the Galaxy and the
SMC.  As more carbon stars were added from the SMC 
\citep{lag07} and LMC \citep{zij06,lei08}, the two tracks in
Figure~\ref{f.csic} remained separate.  \cite{zij06} pointed 
out that metallicity will determine the availability of 
silicon, so that metal-rich stars generally populate the 
sequence with stronger SiC emission while metal-poor stars 
show weaker SiC features.

All eight of the stars with [6.4]$-$[9.3] $\ga$ 2.0 have
negative SiC/continuum ratios, which indicate SiC absorption
in optically thick dust shells.  The seven carbon-rich EROs 
observed by \cite{gru08} all show SiC 
absorption; two of these were re-observed as potential 
post-AGB objects by \cite{mat14}.  The eighth source was part 
of the SAGE-Spec program \citep{kem10}.  All have 
spectroscopic properties consistent with deeply embedded 
objects near the end of their AGB lifetimes.

% IRAS 05042 - cluster021_01 - 40650
% IRAS 05187 (barely) - cluster024_07 - 40650
% IRAS 05191 - cluster028_01 - 40650
% IRAS 05260 - cluster028_03 - 40650
% IRAS 05305 - cluster032_01 - 40650
% IRAS 05026 - 50338 - Matsuura
% IRAS 05509 - 50338 - Matsuura
% IRAS 05133 - 40519 - SAGE-Spec

\subsubsection{Magnesium sulfide dust} % Sec. 4.2.2

\begin{figure} % Fig. 6
\includegraphics[width=3.4in]{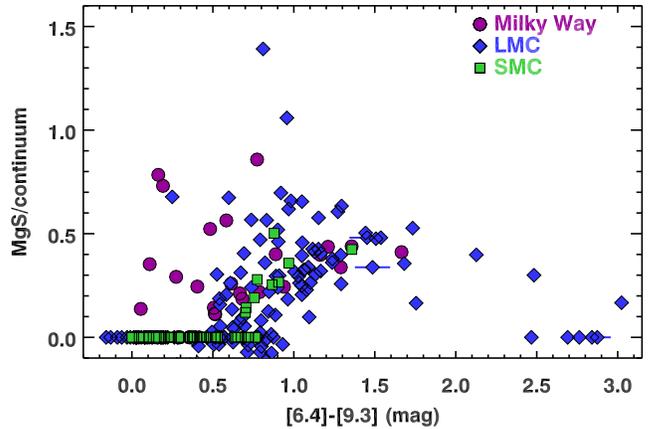} % c69mgs_1511.eps
\caption{The strength of MgS dust emission as a function of 
[6.4]$-$[9.3] color.  More metal-poor stars have to have redder
[6.4]$-$[9.3] colors before MgS becomes visible.\label{f.cmgs}}
\end{figure}

\begin{figure} % Fig. 7
\includegraphics[width=3.4in]{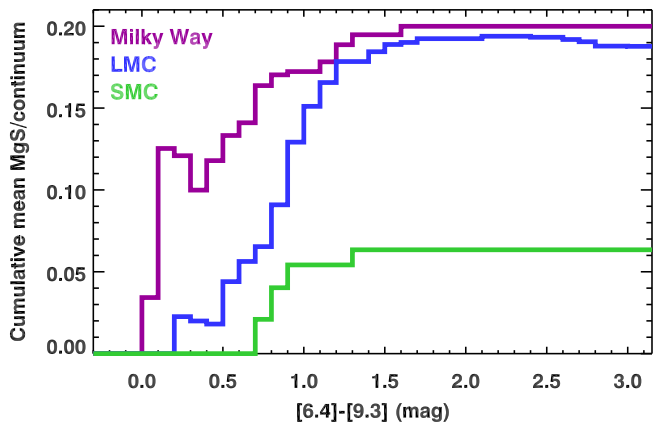} % figmgs2.eps
\caption{The cumulative mean MgS emission.  At each color, 
the mean MgS/continuum strength for all sources with bluer 
colors is plotted.\label{f.mgs2}}
\end{figure}

Figure~\ref{f.cmgs} shows how the MgS emission increases with
[6.4]$-$[9.3] in the different populations.  MgS is apparent
in the Galactic sample at all [6.4]$-$[9.3] colors, while in
the LMC, the first MgS feature is at [6.4]$-$[9.3] $\sim$ 
0.25.  No other LMC source shows MgS until [6.4]$-$[9.3] 
reaches a value $\sim$0.5.  In the SMC, MgS does not appear 
until [6.4]$-$[9.3] $\sim$ 0.7.  

Figure~\ref{f.mgs2} illustrates the behavior of the MgS
emission differently, by plotting the mean MgS/continuum 
strength for all data up to a given [6.4]$-$[9.3] color.  It 
shows more clearly how the LMC lags the Milky Way, with the
SMC even further behind, as the [6.4]$-$[9.3] color 
increases.  The traces for the Milky Way and SMC are
unchanged past colors of $\sim$1.6 and 1.4, respectively, due
to a lack of redder sources.  The lower abundances of Mg and 
S in more metal-poor stars appear to be delaying the 
condensation of MgS until the dust-production rate is higher 
and the dust is cooler.

\subsubsection{Acetylene gas at 7.5~\mum} % Sec. 4.2.3

\begin{figure} % Fig. 8
\includegraphics[width=3.4in]{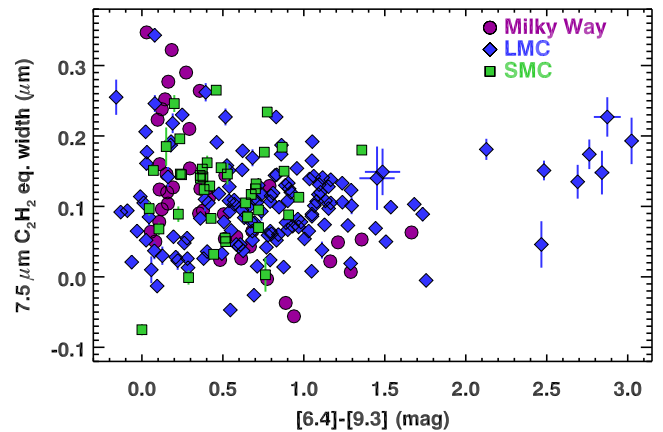} % c69w75_1511.eps
\caption{The strength of the 7.5~\mum\ C$_2$H$_2$ absorption 
band versus [6.4]$-$[9.3] color.\label{f.c2h2}}
\end{figure}

\begin{figure} % Fig. 9
\includegraphics[width=3.4in]{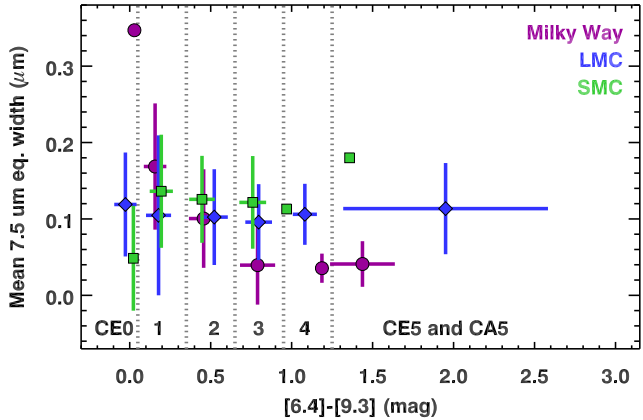} % figgas.eps
\caption{The strength of the 7.5~\mum\ C$_2$H$_2$ absorption
band versus [6.4]$-$[9.3] color, binned by CE class to
illustrate the general trends in the data.  When determining
means, negative equivalent widths have been treated as zero,
because acetylene emission is unlikely.\label{f.gas}}
\end{figure}

Figure~\ref{f.c2h2} compares the behavior of the 7.5~\mum\
absorption band from C$_2$H$_2$ for the different samples.
For the less dusty stars ([6.4]$-$[9.3] $\la$ 0.35), all 
three galaxies show a range of band strengths, but to the 
red, the upper bound diminishes.  Figure~\ref{f.gas} 
clarifies the behavior of each galaxy by plotting the mean 
equivalent width and color for each CE class.  For CE2 and 
redder sources, the mean equivalent width of the 7.5~\mum\ 
acetylene band is consistently strongest in the SMC and 
weakest in the Galaxy.

As shown below (Section~\ref{s.space}), all of the CE0 and 
most of the CE1 sources are associated with relatively
dust-free carbon stars which can have strong molecular
absorption bands due to the lack of veiling from the dust
surrounding the redder sources.  \cite{slo15b} found that
in the SMC, the dusty stars with veiled molecular bands 
usually are pulsating in the fundamental mode with strong
amplitudes as Mira variables, while the other group are
detected as semi-regulars or irregulars.  To focus on the
dust-production process, we should focus on the stars 
classified as CE2 or later.

To make a quantitative comparison of the 7.5~\mum\
acetylene band, we examine the CE2--3 sources, because
the CE4 and CE5 classes have only one SMC source each.
We set negative equivalent widths to zero, because
7.5~\mum\ emission is unlikely.  For the chosen color 
range, the mean equivalent width of the 7.5~\mum\ band 
decreases from 0.12$\pm$0.01~\mum\ in the SMC to 
0.10$\pm$0.01~\mum\ in the LMC and to 0.08$\pm$0.02~\mum\ 
in the Galaxy (the quoted errors are the uncertainty in the 
mean).  While the spread in each sample is substantial, the 
steady drop in mean band strength is still apparent.  Thus 
the expanded IRS sample has not changed the conclusions drawn 
before about the 7.5~\mum\ acetylene band.
% C2H2 analysis - cf notes 23 Sep 15, 

While the behaviors of the C$_2$H$_2$ absorption at 7.5~\mum\ 
and the SiC emission at 11.5~\mum\ differ quantitatively,
their rough similarity raises a possible concern, which we 
have examined and ruled out.  Acetylene also shows a strong 
absorption band centered at 13.7~\mum, but the sharp 
absorption band apparent at 13.7~\mum\ is just from the Q 
branch ($\Delta$$J$ = 0), with the P and R branches extending 
the feature over a micron in either direction.  The possible 
SiC emission could be affected by strong C$_2$H$_2$ 
absorption to the red, which could push the continuum fit 
down and artifically add to the measured emission.  We 
investigated by shifting the wavelengths used to fit the 
continuum on the red side of the SiC feature.  
Figure~\ref{f.mm1} shows an inflection on the SiC feature at 
$\sim$12.0~\mum.  We shifted the red continuum wavelengths to 
this inflection and recalculated the intregated SiC flux for 
our samples, and found that, while the total SiC emission 
certainly grew smaller, the samples behaved the same 
qualitatively, and the differences in how the three galaxies 
behaved were unchanged.

\subsection{Bolometric magnitudes \label{s.mbol}} % Sec. 4.3

\begin{figure} % Fig. 10
\includegraphics[width=3.4in]{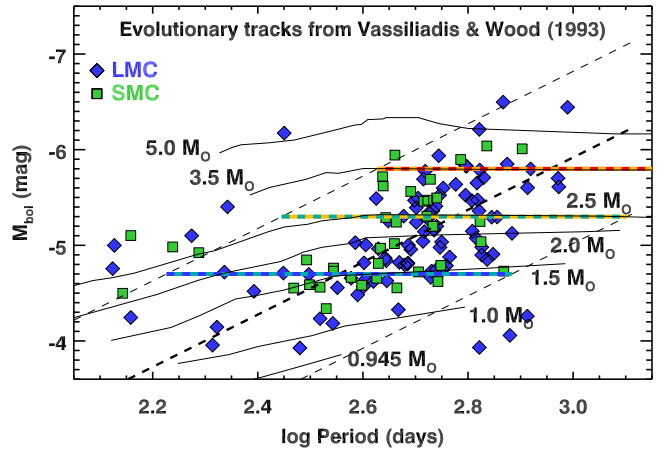} % pl_1603evo.eps
\caption{A period-luminosity plot showing how we binned the
Magellanic carbon stars into four initial-mass groups based 
on their bolometric luminosities.  The thick dashed line is 
the period-luminosity relation fitted to the fundamental-mode 
pulsators, and the thin dashed lines to either side enclose 
the range we will consider to be on the fundamental-mode 
sequence.  Within this range, evolutionary tracks on the AGB 
are generally close to horizontal, as demonstrated by the 
tracks from \cite{vw93}.  The multi-colored horizontal lines 
show the boundaries between the mass bins.\label{f.plevo}}
\end{figure}

For each star in the Magellanic sample, we can estimate its 
bolometric magnitude ($M_{{\rm bol}}$) by integrating its 
spectral energy distribution, which is determined from the 
IRS spectra and the mean magnitudes from the multi-epoch 
photometry described in Section~\ref{s.photo}.  We integrated 
through the photometry below 5~\mum\ using linear 
interpolation, then integrated the IRS spectrum.  To the blue 
of the optical photometry, we assumed a Wien distribution.  
To the red of the IRS data, we assumed a Rayleigh-Jeans tail.  
If the IRS spectrum only included data from SL (i.e.\ stopped 
at 14~\mum), we also utilized the MIPS data at 24~\mum.  The 
resulting estimates for $M_{{\rm bol}}$ appear in the last 
column of Table~\ref{t.mir}.

Our IRS sample includes six carbon stars in NGC~419.  At an 
assumed distance modulus of 18.90, the median absolute
magnitude for these stars = $-$4.79 $\pm$ 0.23, with 
$M_{{\rm bol}}$ = $-$5.24 for NGC~419 IR~1 and $-$4.73 for 
NGC~419 MIR~1.  These values are consistent with an estimated
zero-age main-sequence mass of 1.9~M$_{\odot}$ \citep{kam10}
and an age of 1.35 Gyr \citep{gir09}, based on models of
the AGB and red clump, respectively.  The bolometric
magnitudes of NGC~419 compare well with the rest of the
SMC, which has $<$$M_{{\rm bol}}$$>$ = $-$5.03 $\pm$ 0.49.  
If NGC~419 were in front of the SMC at a distance modulus of 
18.50 \citep{gla08}, then its median $M_{{\rm bol}}$ would 
lie outside the 1$\sigma$ range for the SMC, making it older 
than most of the carbon stars observed by the IRS in the SMC.

Figure~\ref{f.plevo} includes evolutionary tracks from 
\cite{vw93} on the period-luminosity plane and also shows the 
location of the P-L relation for fundamental-mode pulsators.  
For stars with higher initial mass and for lower-mass stars 
with longer periods, the evolutionary tracks are nearly 
horizontal, so that $M_{{\rm bol}}$ can be used as a proxy 
for initial mass.  We have therefore divided the 
fundamental-mode pulsators into four mass bins with the 
horizontal lines defined in Figure~\ref{f.plevo} as 
boundaries.  While the evolutionary tracks do deviate from 
horizontal for shorter pulsation periods, the figure shows 
that this has little impact on the stars in our sample, 
provided we only consider stars with periods $\ga$ 250 days 
(log $P$ $\sim$ 2.4).

It should be emphasized that the estimates for initial mass 
are rough and are not to be taken too literally or 
quantitatively.  The key is the steady bolometric magnitude 
on the late stages of the evolutionary tracks.  These make it 
possible to use $M_{\rm bol}$ to segregate the data by 
initial mass for statistical purposes.  

The thick black dashed line in Figure~\ref{f.plevo} is a
period-luminosity relation, first fitted to all of the 
Magellanic stars in our sample with periods $>$ 250 days, 
then fitted iteratively to only those within an envelope of
$\pm$ 0.9 magnitudes to avoid stars not on the 
fundamental-mode sequence.  The linear solution converges 
quickly:
\begin{equation}
  M_{\rm bol} =  2.30 - 2.74 ~ {\rm log}~P~ {\rm (days)}.
\end{equation}
This line is steeper than the P-L relation by \cite{whi06}.  
They found a slope of $-$2.54 using bolometric corrections to 
estimate $M_{{\rm bol}}$, while our estimates are based on 
integrations through the IRS data and the available 
photometry at shorter wavelengths.  \cite{slo15b} noted that 
the estimated bolometric magnitudes of the dustiest sources, 
which tend to have the longest periods, depend on the precise 
method used.  Here, we have spectral data covering the peak 
of the spectral energy distribution for those sources, which 
should improve on the cases where only photometry is 
available.

\subsection{The period-color relation and initial mass\label{s.cp}} % Sec. 4.4

\begin{figure} % Fig. 11
\includegraphics[width=3.4in]{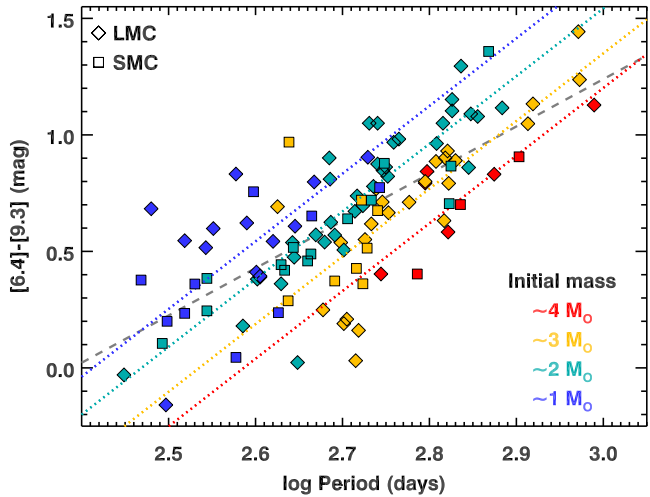} % fpd1_1603.eps
\caption{The [6.4]$-$[9.3] color as a function of pulsation
period with the Magellanic sample, color-coded by their
initial mass as estimated from their bolometric magnitude.
Sources not assigned to a mass bin are not included.
The width of the relation between [6.4]$-$[9.3] color and 
period clearly arises from the mass distribution of the 
samples.  The gray dashed line is a line fitted to all of 
the data, while the colored dotted lines are fitted to
each mass bin separately (see Section~\ref{s.mbol}).
\label{f.pc69m}}
\end{figure}

\begin{deluxetable}{ccr} % Table 7
% cf notes 28 Aug, 28 Nov, 10 Dec 15, 30 Mar 16
\tablecolumns{3}
\tablewidth{0pt}
\tablenum{7}
\tablecaption{Period-corrected differences in [6.4]$-$[9.3] color}
\label{t.coldiff}
\tablehead{
  \colhead{Approximate} & \multicolumn{2}{c}{$\Delta$(LMC) $-$ $\Delta$(SMC) 
  (mag.)\tablenotemark{a}} \\
  \colhead{initial mass (M$_{\odot}$)} & \colhead{Using one fitted line} & 
  \colhead{Fitting to each mass bin}
}
\startdata
   $\sim$4 & 0.083 $\pm$ 0.098 & $-$0.001 $\pm$ 0.077 \\
   $\sim$3 & 0.147 $\pm$ 0.100 &    0.054 $\pm$ 0.103 \\
   $\sim$2 & 0.112 $\pm$ 0.048 &    0.032 $\pm$ 0.043 \\
   $\sim$1 & 0.157 $\pm$ 0.093 &    0.131 $\pm$ 0.100
\enddata
\tablenotetext{a}{$\Delta$ = median \{ ([6.4]$-$[9.3])$_{obs}$ $-$ 
  ([6.4]$-$[9.3])$_{line}$ \}.}
\end{deluxetable}

Figure~\ref{f.pc69m} reprises Figure~\ref{f.pc69}, except 
that the sources are color-coded by our rough estimates of 
their initial mass.  Figure~\ref{f.pc69m} explains the 
width of the relationship of [6.4]$-$[9.3] versus period.  
The Magellanic samples include a range of masses, each of 
which follows its own narrow relation.  When stars switch
from overtone pulsations to the fundamental mode, they
begin to produce dust at significant rates for the first 
time, and this rate increases as their pulsation period
increases.  The slope to the period-luminosity relation means 
that this mode switch occurs at longer pulsation periods for 
more luminous (and thus more massive) stars, which explains 
why segregating by mass reveals the parallel tracks in 
Figure~\ref{f.pc69m}.  The finite width of the 
period-luminosity relation suggests that this scenario of 
increasing pulsation period and increasing DPR and mass-loss 
rate can be sustained for a limited time.

\cite{slo08} concluded that for Magellanic carbon stars, 
metallicity did not play a detectable role in dust 
production, because the Galaxy, LMC, and SMC showed 
indistinguishable relations between [6.4]$-$[9.3] color and 
pulsation period.  However, more metal-poor carbon stars in
other nearby Local Group dwarf galaxies ([Fe/H] $\la$ $-$1)
do show less dust than might be expected for their pulsation
period \citep{slo12}.

The data in Figure~\ref{f.pc69m} point to the need to account 
for the different mass distributions of the carbon stars 
observed by the IRS in the LMC and SMC in order to re-assess 
if the role of metallicity can be detected for [Fe/H] $\ga$ 
$-$1.  A line fitted to all of the data gives
\begin{equation}
   [6.4]-[9.3] = -4.52 + 1.91~ {\rm log}~P~ {\rm (days)}.
\end{equation}
For each star, we used this line to predict the expected
[6.4]$-$[9.3] color based on the pulsation period and
compared it to the observed [6.4]$-$[9.3] color.  For each
mass bin and for each galaxy, we computed the median of these
offsets.  Table~\ref{t.coldiff} gives the difference in
these median offsets, along with the propagated uncertainty
in the mean (in the column labeled ``Using one fitted 
line'').  Averaging the differences between LMC and SMC 
across the mass bins reveals a redder [6.4]$-$[9.3] color in 
the LMC of 0.125 $\pm$ 0.042 magnitudes, which is a 
2.9 $\sigma$ result.

The data in each mass bin in Figure~\ref{f.pc69m} follow a
steeper slope than the total data set, and the carbon stars
in the LMC and SMC do not have the same period distributions.
To further correct for this difference in the sample, we have
fitted lines independently to the four mass bins.  Because
these lines are driven somewhat by the noise in our data, we 
found the weighted mean of the slopes (2.91 mags/dex) and
used this slope for each mass bin.  The colored dotted lines
in the figure illustrate the results.  Table~\ref{t.coldiff} 
gives the resulting median difference in [6.4]$-$[9.3] 
between the LMC and SMC.  The average from the four mass
bins is 0.054 $\pm$ 0.040, which is only a 1.3 $\sigma$ 
result.

We conclude that the full sample of Magellanic carbon stars 
observed by the IRS shows hints of a subtle dependence of 
dust-production rate with metallicity, but once we account 
for different distributions by initial mass and metallicity, 
our sample does not reveal a statistically significant 
result.  The effects of pulsation period and initial mass on 
dust production are much stronger.

\subsection{Color-color and color-magnitude space \label{s.space}} % Sec. 4.5

\begin{figure} % Fig. 12
\includegraphics[width=3.4in]{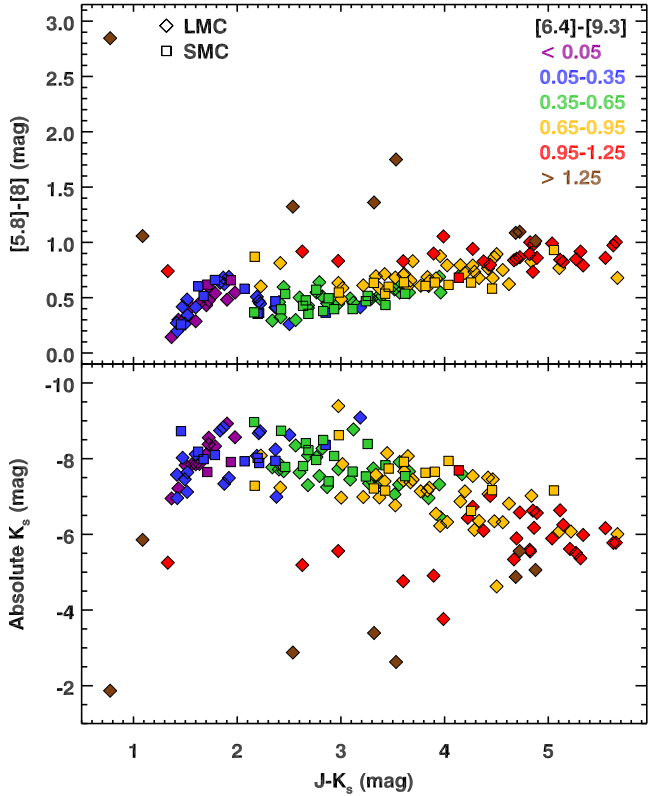} % fccm_jk_1603.eps
\caption{Color-color and color-magnitude diagrams featuring
the $J-K_s$ colors on the horizontal axis.  Symbols are coded
for galaxy with shape and for [6.4]$-$[9.3] color with color.
The [6.4]$-$[9.3] intervals correspond to the CE0--5 
classifications.  None of the CE5s and only about half of the 
CE4s follow the sequences. (CE4:  0.95 $<$ [6.4]$-$[9.3] $<$ 
1.25; CE5:  [6.4]$-$[9.3] $>$ 1.25.)\label{f.ccm_jk}}
\end{figure}

\begin{figure} % Fig. 13
\includegraphics[width=3.4in]{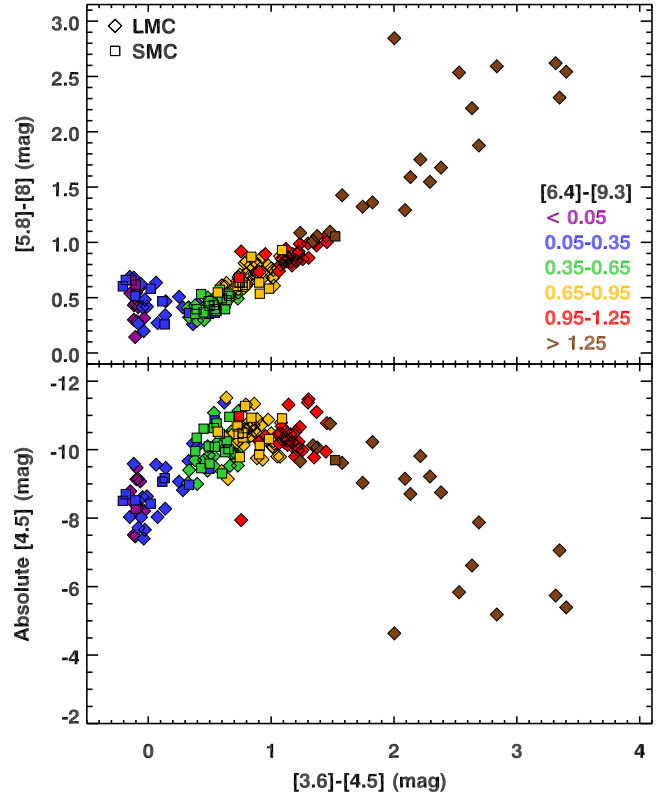} % fccm_lm_1511.eps
\caption{Color-color and color-magnitude diagrams featuring
the $[3.6]-[4.5]$ colors on the horizontal axis.  Symbols are 
coded for galaxy with shape and for [6.4]$-$[9.3] color with 
color.  The [6.4]$-$[9.3] intervals correspond to the CE0--5 
classifications.  Only a handful of the reddest sources in
[5.8]$-$[8] are off the sequence, along with one CE4.
\label{f.ccm_lm}}
\end{figure}

Carbon stars fall along a well-defined sequence in most
infrared color-color diagrams, with each color increasing
steadily and monotonically as the DPR increases.  The 
[5.8]$-$[8] color is an exception, because it is sensitive to 
both dust content and the strength of molecular absorption 
in the 5.8~\mum\ filter \citep{sri11}.  Figure~\ref{f.ccm_jk} 
plots the [5.8]$-$[8] color versus $J-K$ in the top panel.  
From $J-K$ $\sim$ 1.3 to 2, increasing molecular absorption 
reddens [5.8]$-$[8], but for $J-K$ $\ga$ 2, increasing dust 
opacity is the culprit.  \cite{slo15b} called these two 
sequences ``molecular'' and ``dusty'' to identify the
agent responsible for the reddening.  They found that 
semi-regular variables dominate the molecular sequence ($J-K$ 
$\la$ 2), while Mira variables dominate the dusty sequence.  
This difference is not between overtone and fundamental-mode 
pulsators, as half of the semi-regulars are pulsating in the 
fundamental mode.  Rather, it is a difference between stars 
pulsating with small and large amplitudes.  Generally, it is
the latter that are associated with dust production.  

The bottom panel of Figure~\ref{f.ccm_jk} presents the sample
in a more familiar near-IR CMD.  \cite{blu06} labeled the 
sequence extending to the red of $J-K_s$ $\sim$ 2 as the 
``extreme'' AGB stars,\footnote{Formally, their criterion was 
$J$$-$[3.6] $>$ 3.1.} but this terminology is something of a 
misnomer, as the objects are only extreme in the sense that 
they are producing carbon-rich dust, something that all 
carbon stars must do before they end their lives on the AGB.  
Superlatives like ``extreme'' are more appropriate for the 
extremely red objects discovered in the LMC by \cite{gru08}.
These are the deeply embedded carbon stars analogous to 
targets in the Galaxy like AFGL 3068 
\citep[e.g.,][]{lr77,jon78}.  We will follow the practice of 
\cite{slo15b} and refer to objects with $J-K_s$ $\ga$ 2 as 
``dusty'' carbon stars, by which we mean carbon stars 
associated with significant and readily measurable quantities 
of dust.

Figure~\ref{f.ccm_lm} replaces the $J$ and $K_s$ filters 
in Figure~\ref{f.ccm_jk} with [3.6] and [4.5].  In all four
panels of the two plots, the [6.4]$-$[9.3] color generally
increases along the sequence from blue to red, but with some
scatter.  This scatter appears in any color chosen, and it
is consistent with what one would expect from variations in 
DPR as a function of time, with bluer colors sensitive to 
warmer and more recently formed dust.

The CE1 sources (0.05 $<$ [6.4]$-$[9.3] $<$ 0.35) straddle
the boundary between the molecular and dusty sequences in both 
figures.  On the molecular sequence, the overlap between CE0 
and CE1 is complete, despite the presence of more dust in the
CE1 sources, and it shows that the dust is not determining 
position on this sequence.  The vertical spread of the 
molecular sequence in the bottom panel of 
Figure~\ref{f.ccm_lm} shows that the [3.6]$-$[4.5] color is
less sensitive to molecular bands than $J-K_s$.

In Figure~\ref{f.ccm_jk}, none of the CE5 stars 
([6.4]$-$[9.3] $>$ 1.25) follow the relationship between
[5.8]$-$[8] and $J-K_s$ established by the sources with less
dust.  If they did, they would be off the right-hand edge of
the plot, provided we could detect them at $J$ and $K_s$.
Because these sources are so embedded, if they followed the
dusty sequence, they would actually be below the detection 
limit of the near-IR surveys we examined.  Of the 21 CE5 
sources, we have valid $J-K_s$ colors for only nine, and all 
appear with bluer $J-K_s$ colors than the dusty sequence 
would lead us to expect.  In the bottom panel, all of the CE5 
sources plotted are also off the sequence.  Their behavior 
is fully consistent with a shift to bluer $J-K_s$ colors.  
Several of the CE4 and even some CE3 sources can also be seen 
to have shifted.  In Figure~\ref{f.ccm_lm}, no sources are 
missing, and the number of off-sequence sources is greatly 
reduced, with only a few CE5 sources not on the dusty 
sequence in either panel.  

As noted in Section~\ref{s.photo}, it is only for the deeply
embedded sources that a photometric mismatch is a danger, and 
even then the odds of a mismatch are slight for a given 
source.  We conclude that most or all of our off-sequence 
sources have colors indicative of the source itself and not a 
chance superposition of two sources in the search beam.

\subsection{Variability \label{s.var}} % Sec. 4.6

\begin{figure} % Fig. 14
\includegraphics[width=3.4in]{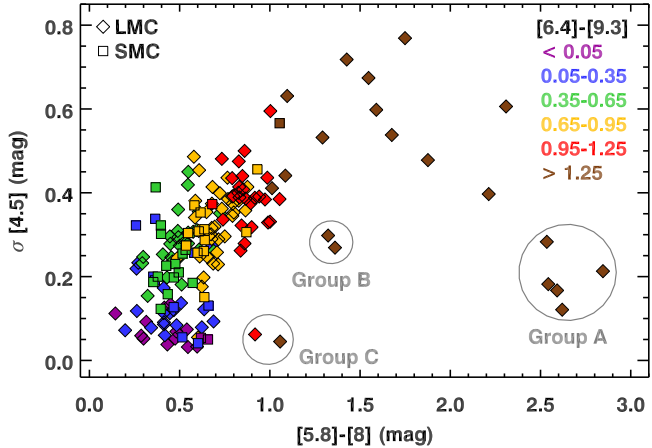} % figvar.eps
\caption{The amplitude of the variability at 4.5~\mum\ (as
measured by the standard deviation of the IRAC and {\it WISE} 
data) as a function of [5.8]$-$[8] color.  The amplitude 
increases steadily from the naked carbon stars to the CE4 
sources, then drops for those with the reddest [5.8]$-$[8] 
and [6.4]$-$[9.3] colors.  The circled groups are discussed
in Section~\ref{s.var}.\label{f.var}}
\end{figure}

If the off-sequence sources in Figure~\ref{f.ccm_lm} have
really moved off the AGB, then the amplitude of their 
variability should have dropped.  Figure~\ref{f.var} plots 
the variability amplitude of all of our carbon stars as a 
function of [5.8]$-$[8] color.  The values plotted are the 
standard deviations ($\sigma$) of the mean magnitudes in 
Tables~\ref{t.optnir} and \ref{t.mir}.  For a pulsation cycle 
which follows a sine function, the peak-to-peak amplitude =
$2 \sqrt{2} \sigma$.

As the stars shift from CE0 (nearly naked) to CE5 (very 
dusty), the amplitude of their variability increases as they 
grow redder, but once [5.8]$-$[8] reaches a value of 
$\sim$1.5, the amplitude begins to decline.  The three 
sources furthest from the carbon sequence in the top panel of 
Figure~\ref{f.ccm_lm} are all in Group~A in the bottom
right-hand corner of Figure~\ref{f.var}.  These are both the
reddest and the least variable sources in the sample.  More 
recent photometry of the reddest carbon stars in the LMC at 
3.6 and 4.5~\mum\ extends the temporal baseline and confirms 
the trend of decreasing variability with increasing 
[5.8]$-$[8] color (Sargent et al., in preparation).

Had we chosen to use the variability amplitude at 3.6~\mum\ 
instead of 4.5~\mum, the results would have been the same 
qualitatively.  Dropping the color corrections derived in 
Appendix~A also does not change the result qualitatively.  
It does change the apparent variabilities quantitatively, 
increasing the apparent amplitudes for the stars which are 
most variable because it increases the apparent scatter in 
the data, but for the relatively non-variable sources, this 
effect is much smaller.  By using the 4.5~\mum\ data, we have 
reduced possible errors from the corrections even more, since 
the {\it WISE}-to-IRAC corrections are much smaller at 
4.5~\mum\ than at 3.6~\mum.

Four more sources appear to be relatively non-variable for
their [5.8]$-$[8] color, and are labelled as Groups B and
C in Figure~\ref{f.var}.  Group C includes one CE4 source
(red diamond), SAGE J054546, which is also off-sequence in 
the bottom panel of Figure~\ref{f.ccm_lm}.  The remaining 
three are all CE5, and except for unusually low-variability 
amplitude, none stands out in any other significant way.

\section{Discussion \label{s.discuss}} % Sec. 5.0

\subsection{Evolution off the AGB} % Sec. 5.1

The three sources most clearly off the carbon sequence in the 
color-color diagram in Figure~\ref{f.ccm_lm} have [5.8]$-$[8]
$>$ 2.5 and [3.6]$-$[4.5] $<$ 3.  They are IRAS~05133, 
IRAS~05191, and IRAS~05260, in order of increasing 
[3.6]$-$[4.5] color.  The other two sources with [5.8]$-$[8] 
$>$ 2.5 are IRAS~05026 and IRAS~05042.  Group A in 
Figure~\ref{f.var} consists of these five stars.  They are
much less variable than the rest of the sources with 
[5.8]$-$[8] $\ga$ 1.5.  And all five are also SiC absorption 
(CA) sources.

In the top panel of Figure~\ref{f.ccm_jk}, the only CA source
with a $J-K_s$ color is IRAS~05133, and it is bluer than any 
other carbon star in our sample!  Whatever has led to its
unusually blue [3.6]$-$[4.5] color has allowed us to detect
an even bluer source at $J$ and $K_s$.  With $K_s$ = 16.6
and $J$ = 17.4, it is just at the detection limit.  The other 
four sources are redder at [3.6]$-$[4.5], have more 
extinction in the near-IR, and consequently are undetected at 
$J$ and $K_s$.

Thus the five most embedded carbon stars in our sample (as 
measured with [5.8]$-$[8]) have properties consistent with 
evolution off the AGB.  Their variability is lower than the 
less embedded sources, they show SiC absorption, indicating 
high column densities of dust, and three appear to be 
developing double-peaked SEDs as the central star becomes 
visible.  

However, other stars are relatively non-variable, and others 
also show SiC absorption.  Of the other non-variable sources, 
all but SAGE J054546 (CE4) are either CE5 or CA5 (i.e.\ 
[6.4]$-$[9.3] $>$ 1.25).  These sources may be approaching 
the end of their AGB evolution.  

The steady decline in variability past [5.8]$-$[8] $\sim$ 1.5 
(Figure~\ref{f.var}) suggests that more than just the handful 
of sources most obviously off of the sequences in 
Figure~\ref{f.ccm_lm} are approaching the end of their AGB
lifetimes.

\subsection{Non-spherical dust shells} % Sec. 5.2

For the off-sequence sources, the excess blue flux indicates
that we are able to peer into the dust and glimpse the 
central star, either directly or through scattered light.  
The dust envelope may be growing patchy or it may be 
distributed asymmetrically, perhaps in a torus or possibly 
even a disk.  

Figure~\ref{f.ccm_jk} reveals other sources which do not 
follow the carbon-star sequence at $J$ and $K_s$, even though 
they are on the sequence in the longer-wavelength filters in 
Figure~\ref{f.ccm_lm}.  In particular, a number of CE4 
sources (in red) are off-sequence in both panels of 
Figure~\ref{f.ccm_jk}, with much bluer $J-K_s$ colors than
expected for either their [5.8]$-$[8] colors or $K_s$ 
magnitudes.  More of the CE5 sources also behave similarly.

The opacity of amorphous carbon dust drops as 
$\sim$$\lambda^{-2}$, making it difficult to explain how we 
could directly observe the central source at $J$ and $K$ but 
not at 3.6 and 4.5~\mum.  If we are detecting the central 
source via light scattered above and below a disk or in the 
polar regions of an asymmetric dust shell, then we would 
expect to see a bluer color at $J-K_s$ than at [3.6]$-$[4.5], 
because the scattering efficiency drops as $\lambda^{-4}$.
\cite{jbs06} proposed scattering in a system with a disk
viewed close to edge-on to explain the unusual object
SMP~LMC~011, which presents an optically thick carbon-rich
dust spectrum with molecular absorption bands in the infrared 
but shows H$\alpha$ in emission in the optical.  \cite{slo95} 
suggested scattering from the poles of an asymmetric dust 
distribution in the embedded Galactic carbon star IRC +10216 
to explain emission from relatively warm dust above and below 
the central source.  This hypothesis could be tested for the 
embedded Magellanic carbon stars by measuring their 
polarization as a function of wavelength in the near-IR.

The nature of the dust geometry has significant consequences.
\cite{slo95} noted that the deviation from spherical
symmetry they argued for in IRC +10216 was subtle and did not 
prevent spherically symmetric radiative transfer models from 
working.  Disks are a different story.  Their non-isotropic 
emission would lead to significant uncertainties in 
bolometric luminosities and thus our basic knowledge about 
the sources.  \cite{boy12} noted that the ten dustiest AGB 
stars detected in the SAGE surveys of the SMC accounted for 
17\% of the total dust input to the galaxy when assuming
spherical symmetry.  If the dust in embedded carbon stars 
were in disks and not outflowing shells, the amount of dust 
produced by AGB stars would have to be revised downward,
because the dust-production rate scales with outflow 
velocity, which in disks would be zero. 

Disks imply binarity, which leads to more consequences.  
Equatorial disks can result from binaries interacting during
the AGB phase \citep[e.g.,][]{sl89}.  If the sources we have 
identified as off-sequence are interacting binaries, then 
they may have have never been on sequence and are instead
following a different evolutionary path in color-color space.
High-resolution imaging of Galactic sources show that 
binary interactions are not rare at all.  The Mira variable
R Scl is embedded in a spiral structure indicative of 
interactions of the outflows from the carbon star with a 
companion \citep{mae12}.  L$_2$ Pup, the nearest AGB star in
the sky, has a dusty disk, evidence of a companion, and the
beginnings of bipolar nebulosity \citep{ker14,ker15}.  These 
observations add to the evidence that binaries and their
interactions on the AGB strongly influence the structure of 
post-AGB objects and planetary nebulae 
\citep[e.g.,][]{lc14,dj15}.

As already stated, near-IR polarimetry of the most embedded
carbon stars in the present sample would help diagnose the 
geometry of these systems.  Radiative transfer modeling would 
also help by translating the constraints imposed by the SEDs 
to constraints on the dust morphology.

\subsection{SiC and luminosity} % Sec. 5.3

Carbon stars in the SMC generally show very little SiC 
emission compared to their counterparts in the LMC and the
Galaxy, but there are exceptions.  \cite{slo06} found five
stars in the SMC that had significantly more SiC for their 
[6.4]$-$[9.3] color, and the present sample adds two more
SiC-rich sources, both with bluer [6.4]$-$[9.3] colors than
the other five (see Figure~\ref{f.csic}).  \cite{slo06} 
examined a number of properties for the five SiC-rich stars 
in their sample and could not find anything that 
distinguished them besides their SiC emission strength.

To compare the SiC-rich and SiC-poor sources in the current
sample, we concentrate on [6.4]$-$[9.3] colors between 0.35
and 0.95 (CE2--3), where the two tracks are most distinct,
and consider the sources above and below SiC/continuum = 
0.125 separately.  In the SMC, the mean bolometric magnitude
for the SiC-rich group is $-$5.05 $\pm$ 0.41, compared to
$-$5.08 $\pm$ 0.50 for the SiC-weak group.  The uncertainties 
in the mean are 0.17 and 0.11, respectively.  The two groups 
are indistinguishable in luminosity, which implies that they
have similar mass, age, and metallicity distributions.

In the LMC, the SiC-rich group has $M_{{\rm bol}}$ = $-$4.91 
$\pm$ 0.45, versus $-$5.20 $\pm$ 0.53 for the SiC-poor group.  
The uncertainties in the mean are 0.08 and 0.09, 
respectively.  While the separation is larger and more 
statistically significant, it is in the opposite sense to 
what we might expect, with the more luminous stars, which 
presumably would be more massive and more metal-rich, 
associated with the weaker SiC features.
% SiC analysis - cf notes 15 Aug 15

Thus, while the relative numbers of stars showing strong
and weak SiC emission in the Galaxy, LMC, and SMC support
the hypothesis that the strength of the SiC feature is
tracing metallicity, we find no support for the hypothesis
{\it within} a given galaxy.  \cite{slo06} were unable to
explain why some sources in the SMC showed strong SiC 
emission while most did not, and we are unable to improve
on that situation.

\subsection{Layered grains \label{s.layer}} % Sec. 5.4

\begin{figure} % Fig. 15
\includegraphics[width=3.4in]{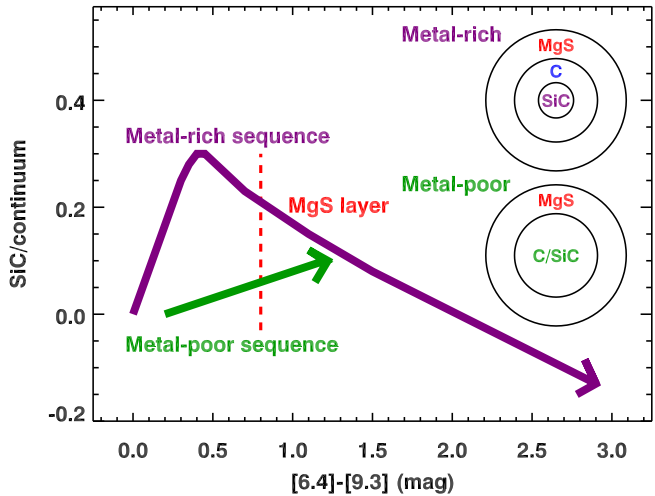} % sictrack.eps
\caption{A possible scenario for the evolution of dust 
grains, modeled on a figure by \cite{lei08}.  The dust around 
more metal-rich stars tends to follow the upper track, with a 
rapid growth in SiC emission with increasing [6.4]$-$[9.3] 
color, then a steady decline until the feature finally goes 
into self-absorption in the most optically thick shells.  
Around more metal-poor stars, the SiC emission grows much 
more gradually as the dust shells redden.  MgS emission 
begins at different [6.4]$-$[9.3] colors, depending on the
metallicity of the star.\label{f.sictrack}}
\end{figure}

\cite{lag07} suggested that the condensation sequence of SiC 
and amorphous carbon differed in metal-rich and metal-poor 
stars, with SiC forming first in the Galaxy and later in the
SMC.  \cite{lei08} followed with a proposal for multiple
evolutionary tracks for dust grains, with the extremes
determined by whether SiC or amorphous carbon formed first.
Figure~\ref{f.sictrack} simplifies the proposal of 
\cite{lei08} and presents two tracks consistent with the
infrared properties of Galactic and Magellanic carbon stars.
The key is that the dust grains must be layered, which may
wreak havoc with attempts to use the grain properties to
estimate the masses of the refractory elements in dust 
grains.

MgS is a case in point.  MgS grains do not form until the dust 
shells pass an optical thickness limit which depends on the
metallicity of the sample, as Figures~\ref{f.cmgs} and
\ref{f.mgs2} show.  The MgS could condense as grains 
independently, but then, in order to explain the strength of 
the MgS feature in post-AGB objects and young planetary 
nebulae (PNe), more S would be required than can exist around 
these stars.  On these grounds, \cite{zha09} argued that MgS 
could not be the carrier of the 26--30~\mum\ feature.

\cite{zij06} proposed that MgS formed as a layer on 
pre-existing seeds of SiC and amorphous carbon.  \cite{lom12} 
argued that the optical properties of a grain coated with MgS
would mimic those of a solid MgS grain, which would account
for the strength of the MgS feature in PNe without violating
abundance constraints.
\cite{slo14} explained why other concerns raised about MgS as 
the carrier of the 26--30~\mum\ feature can be discarded.  
All of the evidence is consistent with MgS forming a layer on 
carbonaceous grains, with the condensation trigger depending 
on metallicity. 

However, it is unlikely that any of the layers are pure.  The
metal-rich track suggests that metal-rich carbonaceous grains
should consist of SiC cores surrounded by amorphous carbon
mantles.  In the most optically thick dust shells, these 
would be mostly coated by MgS, and yet the spectra from these 
sources show {\it absorption} from SiC.  Thus some SiC must 
be present within one optical depth of the surface of the 
grains,  which means that not all grains are coated with MgS, 
or the coatings are incomplete.  Furthermore, if the grains
consist of a SiC core and an amorphous carbon mantle, that
mantle cannot be pure and must also contain some SiC.

Young carbon-rich PNe in the LMC and SMC often show unusually 
strong emission features at $\sim$11.5~\mum\ in their spectra 
\citep{jbs09}.  \cite{slo14} showed that the shape of these 
features was consistent with SiC dust, modified by the 
presence of emission from polycyclic aromatic hydrocarbons 
(PAHs).  From abundance arguments, one would expect weaker 
SiC emission in more metal-poor objects, but the opposite is 
true, with carbon-rich Magellanic PNe showing stronger SiC 
emission than in the Galaxy.  \cite{slo14} suggested that 
photoprocessing of carbonaceous grains would preferrentially 
remove amorphous carbon from the surface, resulting in grains 
coated by SiC.  Such a scenario would require some 
intermixing of SiC and amorphous carbon in the outer layers 
of the grains.

The available observational evidence points to layered grains 
with a structure more complicated than depicted in 
Figure~\ref{f.sictrack}.  The strength of the SiC and MgS
features with increasing [6.4]$-$[9.3] color points to two
tracks as shown, but the presence of SiC absorption in the
most embedded sources and the strength of the SiC feature in
young PNe requires some SiC in the layers of the grains 
within one optical depth of the surface.

\subsection{Completeness} % Sec. 5.5

\begin{figure} % Fig. 16
\includegraphics[width=3.4in]{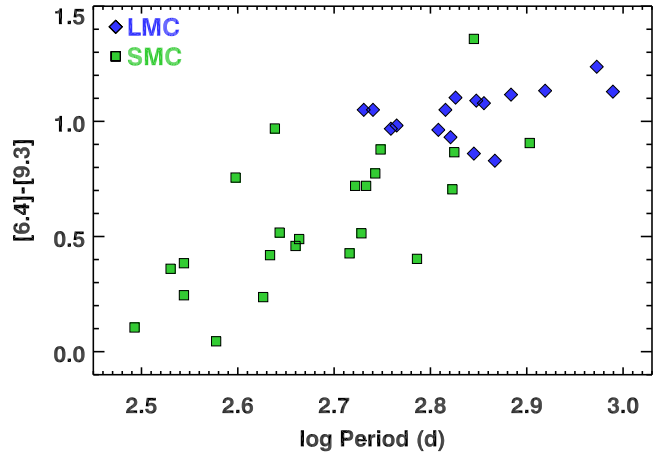} % fcompvl.eps
\caption{The samples of Magellanic carbon stars studied by
\cite{vl06, vl08} plotted on the [6.4]$-$[9.3] versus
period plane.  The SMC sample shows less dust than the LMC
sample because it generally probes a sample with lower 
pulsation periods.  Both samples follow the same trend of
increasing dust content with increasing pulsation period.
\label{f.vl}}
\end{figure}

The pointed nature of spectroscopic samples introduces 
biases, which can make it challenging to compare different
populations.  This work has focused on using properties like 
the pulsation period of the star or the [6.4]$-$[9.3] color, 
which is a proxy for the amount of dust, as the independent 
variable in order to compare sources with similar properties
when examining other properties such as the strength of 
emission features from SiC or MgS dust.  A key result is that 
the amount of dust produced by a star in the Galaxy, LMC, or 
SMC, depends primarily on its pulsation period and initial 
mass.  The effect of metallicity is unclear with the current 
sample.  While we do see hints of less dust in more 
metal-poor stars when controlling for period and mass, this 
decrease is not statistically significant in the IRS samples.

In contrast, \cite{vl08} found a clearer difference in the 
dust content of their sample of carbon stars in the
SMC, compared to a similar study in the LMC \citep{vl06}.
They drew this conclusion from a spectroscopic study of the 
absorption bands from C$_2$H$_2$ at 3.1 and 3.8~\mum,
finding that dust veiling affected the bands more in the
LMC than in the SMC.  Figure~\ref{f.vl} plots the 
[6.4]$-$[9.3] color for the stars in common between their 
sample and ours as a function of pulsation period and shows
that while the carbon stars in the LMC have more dust on
average than their counterparts in the SMC, the two samples 
have different distributions in pulsation period, as well.  
None of the carbon stars in the LMC in common with the IRS 
sample have a period less than 500 days, compared to over 
half the SMC sample.  Given the differences in the LMC and 
SMC samples, one would expect the stars in the LMC to show 
more dust, simply because they are pulsating with longer 
periods.

The IRS-based studies have concentrated on comparing like 
with like, which means comparing the dust content in stars 
with similar pulsation periods.  Previous studies with the
IRS on {\it Spitzer} have found little dependence of the
dust content on metallicity in carbon stars.  The larger
sample considered here has not changed that result.

However, the evidence is building that the populations of
carbon stars in LMC and SMC differ.  Figure~\ref{f.vl} shows 
that the 3~\mum\ samples contain more longer-period pulsators 
and dustier carbon stars in the LMC than the SMC.  The IRS 
samples show a similar difference in the two galaxies.  In 
the SMC sample, the reddest [6.4]$-$[9.3] color is 1.4, so 
that only one SMC source out of 40, NGC~419 MIR~1, is 
classified as CE5, compared to 20 CE5 and CA5 sources out
of 144 in the LMC, 15 of which are redder than NGC~419 MIR~1, 
with [6.4]$-$[9.3] colors extending past 3.0.  At first 
glance, this could easily be a selection effect, as is often 
the case for spectroscopic samples, since the carbon-rich 
component of at least three IRS programs focused exclusively 
on the most embedded objects in the LMC.  However, analogous 
objects do not appear to be present in the SMC.  Our searches 
for deeply embedded carbon stars in the SMC using photometry 
from {\it Spitzer} and {\it WISE} have been unsuccessful.  
Searches using mid-IR photometry from {\it Spitzer} confirm 
this result \citep{del15,sri16}, and searches in the 
far-IR with {\it Herschel} have also not uncovered any deeply 
embedded AGB stars in the SMC \citep{jon15}.  These missing 
objects are likely to be massive carbon stars not present in 
the SMC due to a lower rate of recent star formation 
\citep[e.g.][]{ven16}.

This deficiency of embedded stars in the SMC combined with 
how these sources move off the carbon sequence in near-IR 
color-magnitude space solves a problem first noticed by 
\cite{lag10}.  They derived a relation between absolute 
magnitude at $K_s$ and $J-K_s$ as a means of estimating the 
distances to carbon stars, but they found much more scatter 
in the LMC sample than the SMC and used the latter galaxy to 
derive a relation.  Figure~\ref{f.ccm_jk} shows that  the 
off-sequence embedded carbon stars that appear only in the 
LMC sample are responsible for this apparent scatter.  This 
new finding presents a new problem for the use of 
color-magnitude relations to find the distance to carbon
stars, because a given star might be off sequence.  A 
possible solution would be to develop and use 
color-magnitude relationships at multiple wavelengths to 
estimate distance and to remove outliers at shorter 
wavelengths.  Alternatively, off-sequence stars could
be identified from their reduced variability.

\subsection{The dust-production trigger} % Sec. 5.6

The dichotomy between the carbon stars on the molecular
and dust sequences described by \cite{slo15b} points to
pulsation as the trigger for dust production.  The carbon
stars producing dust are almost entirely Mira variables,
while the carbon stars on the molecular sequence are
either pulsating in one or more overtones, or they are
pulsating weakly in the fundamental mode.  The link 
between strong pulsation and carbon-rich dust production
is clear, even if which is the cause and which is the
effect is not.

Abundance arguments suggest that dredge-up is the trigger.  
As described in the introduction, similar quantities of 
dredged-up carbon will lead to a higher amount of free carbon 
in a more metal-poor star \citep{mat05,slo12}.  The DPR 
should be related to the free carbon abundance, and yet we 
clearly do not observe higher DPRs in more metal-poor carbon 
stars.  \cite{slo12} suggested that once the free carbon 
abundance crossed a critical threshold, dust production would 
spike, stripping the stellar envelope and ending its 
evolution on the AGB.  \cite{lz08} argued similarly, 
suggesting that the final dredge-up on the AGB would initiate 
a superwind.  Crossing a free-carbon threshold is the more 
likely trigger.

Dredge-ups affect atmospheric opacity and generate more
consequences than just free-carbon abundance.  Higher 
opacities will increase the stellar radius, lowering the
escape velocity and making dust-production and mass-loss 
easier.  An expanded and more opaque atmosphere may also be 
more unstable to pulsations, which would in turn push
material out at velocities closer to the now lowered escape
velocity.  In other words, the higher pulsation amplitudes
and increased dust production may both result from the same
underlying cause, the last critical dredge-up on the AGB.

What is intriguing about the sample of IRS-observed 
Magellanic carbon stars is that it may include some sources
on the other side of that final superwind.  Because of the
selection biases that defined this sample, it cannot provide
meaningful relative statistics of carbon stars in different 
phases of their lifetimes on the AGB.  But it can inform the 
photometric studies needed to study these fundamentally
important phases of stellar evolution.

\section{Summary \label{s.sum}} % Sec. 6.0

The IRS on the {\it Spitzer Space Telescope} obtained spectra
of 144 carbon-rich AGB stars in the LMC and 40 in the SMC.
Comparison to Galactic samples reveals no strong dependence 
of the amount of dust produced with metallicity.  Using the 
bolometric magnitudes to estimate initial mass and accounting
for the different mass distributions hints at a subtle
decrease in dust-production rate from the LMC to the SMC,
but the result is not statistically significant 
(1.3 $\sigma$).

The amount of SiC dust accompanying the amorphous carbon
does depend on metallicity, with strong SiC features observed
in most of the Galactic sample and roughly half of the LMC 
sample.  Most of the carbon stars in the SMC show much
weaker SiC emission.  Some of the SMC stars are exceptions
and show stronger SiC emission, but the reason is unknown.
Generally, as the optical depth of the dust increases, the
SiC emission rises to a peak and then shifts into 
self-absorption.  The LMC is unique among the three samples
considered here in that several sources show the SiC feature
strongly in absorption.

The strength of the MgS feature at $\sim$30~\mum\ also
depends on metallicity, with lower metallicity leading to the
appearance of the feature in more optically thick dust 
shells.  This behavior is consistent with previous 
arguments that the MgS condenses as a layer onto pre-existing
grains.

Photometry in the near-IR and at 3.6 and 4.5~\mum\ reveals
some radiation from the central stars in many of the more 
embedded carbon stars.  This excess blue emission shifts 
these sources off the usual carbon sequence seen in most 
infrared color-color and color-magnitude diagrams.  The 
photometric behavior of these off-sequence stars is 
consistent with either a patchy dust shell or scattered light 
in a system with less dust above its poles, possibly a disk.

The pulsation amplitudes generally increase as the dust
grows optically thick.  However, the most embedded sources
in our sample are relatively non-variable, show SiC
absorption in their spectra, and also show signs that they
are revealing their central star.  This behavior is
consistent with evolution off the AGB.  Their observed
properties are also consistent with interacting binary
systems, which complicates their interpretation.

\section*{Appendix~A --- Photometric relations for carbon 
stars} % Sec. App. A

\begin{figure} % Fig. 17
\includegraphics[width=3.4in]{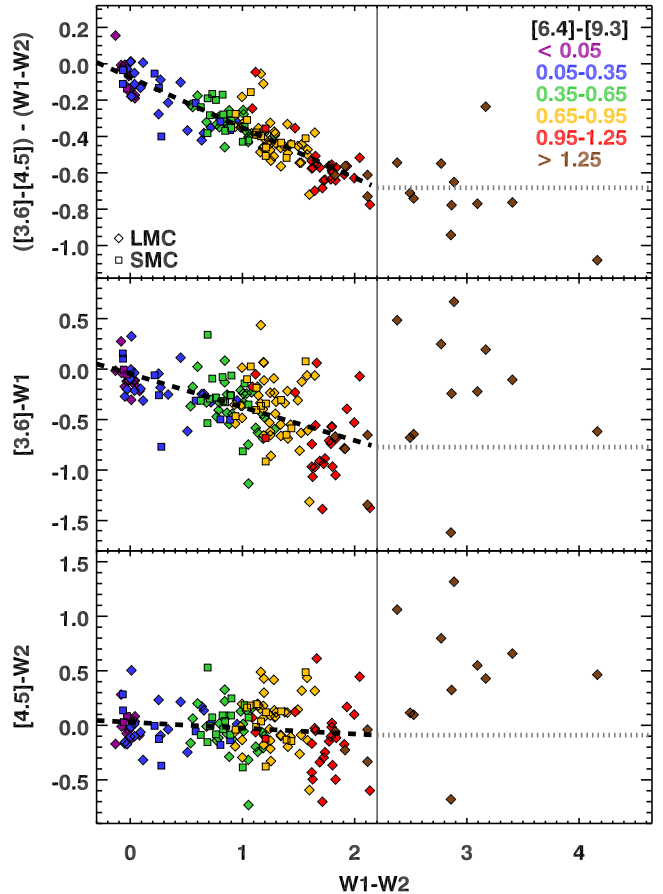} % figcolcorr.eps
\caption{Relations between IRAC and {\it WISE} colors and 
magnitudes, as a function of W1$-$W2 color, for carbon
stars.  Lines are fitted to the data with W1$-$W2 $<$ 
2.2.\label{f.corr}}
\end{figure}

\begin{figure} % Fig. 18
\includegraphics[width=3.4in]{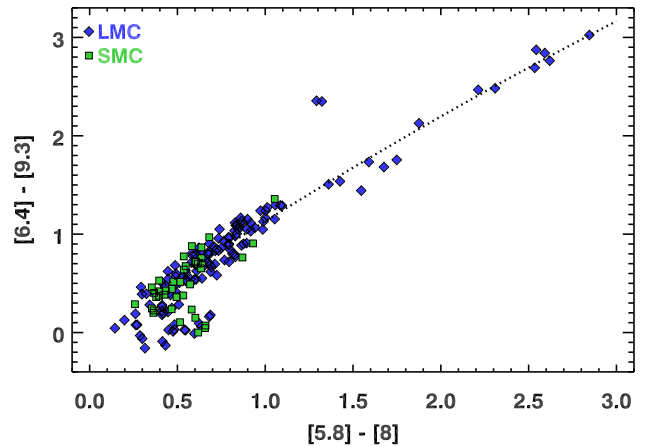} % cc588_69.eps
\caption{The relationship between [6.4]$-$[9.3] and 
[5.8]$-$[8] colors and a quadratic fitted to data with
[6.4]$-$[9.3] $>$ 0.35.  The two data points at [6.4]$-$[9.3] 
$\sim$ 2.3 and [5.8]$-$[8] $\sim$ 1.3 are badly mispointed, 
resulting in artificially red [6.4]$-$[9.3] colors, which 
have been replaced in the data tables using estimates from 
the fitted relation (see text).
\label{f.cc}}
\end{figure}

\begin{deluxetable}{lrr} % Table 8 - cf notes 20 Nov 15 p 3
\tablecolumns{3}
\tablewidth{0pt}
\tablenum{8}
\tablecaption{Color and magnitude corrections for carbon stars}
\label{t.corr}
\tablehead{ \colhead{Relation} & \colhead{y-intercept} & \colhead{Slope} }
\startdata
  ([3.6]$-$[4.5]) $-$ (W1$-$W2) versus W1$-$W2 & $-$0.0752 & $-$0.2759 \\
  {[}3.6] $-$ W1 versus W1$-$W2                & $-$0.0469 & $-$0.3298 \\
  {[}4.5] $-$ W2 versus W1$-$W2                &    0.0283 & $-$0.0539
\enddata
\tablecomments{These corrections are only valid for W1$-$W2 $<$ 2.2.}
\end{deluxetable}

The repeated infrared photometric surveys of the Magellanic
Clouds provide multiple epochs in the IRAC and {\it WISE} 
filter systems.  Because the {\it WISE} filters at 3.4 and 
4.6~\mum\ overlap the IRAC filters at 3.6 and 4.5~\mum, it is 
possible to develop color-based transformations so that we 
can convert the {\it WISE} data to the IRAC system.  
The result is up to ten or more epochs over baselines of 
several years for some of the targets in our sample.

Figure~\ref{f.corr} plots the differences between the IRAC
and {\it WISE} systems as a function of the W1$-$W2 color 
([3.4]$-$[4.6]).  The top panel shows that for carbon stars, 
the difference between IRAC [3.4]$-$[4.5] and {\it WISE} 
colors is well behaved.  We could fit the entire sequence 
with a quadratic, but instead fit a line for W1$-$W2 $<$ 2.2.  

The lower two panels in Figure~\ref{f.corr} illustrate the 
need for a cut-off at W1$-$W2 = 2.2.  To the blue, the
relations of [3.6] to W1 and [4.5] to W2 are clear enough,
although with significant noise due to our limited sampling
and the variability of the sources.  This variability is
suppressed in the top panel because the color changes much
less than the magnitudes over the pulsation cycle.  In the 
lower two panels, however, the data to the red of W1$-$W2 = 
2.2 do not conform to the behavior of the bluer sources, with 
all but one data point falling above any line fitted to the 
blue data.  An examination of Figure~\ref{f.var} hints at 
what must be happening.  All of the stars with W1$-$W2 $>$ 
2.2 are classified as CE5 or CA5.  From CE0 to CE4, the 
pulsation amplitude increases steadily, but most of the CE5s 
are on the other side of the peak amplitude.  These stars are 
likely to be near the end of their AGB lifetimes and may be 
exhibiting non-spherical dust geometries.  For lack of a 
better solution for the redder sources, we will assume that 
the corrections are flat past W1$-$W2 = 2.2 when combining 
{\it WISE} and IRAC data to determine mean magnitudes at 3.6 
and 4.5~\mum.

Because some sources are missing data in the different
filters involved, the corrections determined for individual 
magnitudes do not quite produce the correction for color 
when combined algebraically.  We have modified the magnitude
corrections slightly so that they do.  Table~\ref{t.corr} 
gives the color-based corrections for all three panels in 
Figure~\ref{f.corr}.  The shift between [4.5] and W2 is 
flatter than between [3.6] and W1 because the 
longer-wavelength filters overlap more.  

Figure~\ref{f.cc} illustrates the relationship between the
IRAC [5.8]$-$[8] color and the [6.4]$-$[9.3] color derived
from the IRS spectra.  We excluded the two outliers with
[5.8]$-$[8] $\sim$ 1.3 when fitting a quadratic to the
data.  The polynomial coefficients are $-$0.0282, 1.2074, 
$-$0.0476.\footnote{Polynomical coefficients are listed with
the y-interecept first.}  The standard deviation of the 
data about this function is 0.1085.
% cf notes 23 Nov 15

The two outliers are IRAS~04589 and IRAS~05306, both from 
Program 37088.  The IRS observations were centered on a 
nearby 2MASS star and were somewhat mispointed for the 
intended target, leading to a loss of 75--80\% of the flux in 
SL compared to LL.  As a result, the targets did not appear 
as point sources in the spectral images, the optimal 
extraction broke down, and we used the tapered-column 
extraction method for these sources.  The truncation of 
starlight by the spectroscopic slit in such a case is a 
function of wavelength, leading to an artifically reddened 
spectrum and a false measurement of the [6.4]$-$[9.3] color.  
For this reason, we replaced the [6.4]$-$[9.3] color for 
these two sources in all data tables and figures with values 
estimated from their [5.8]$-$[8] color.\footnote{IRAS~04589 
and IRAS~05306 have [5.8]$-$[8] colors of 1.292 and 1.323, 
respectively, giving estimated [6.4]$-$[9.3] colors of 1.452 
and 1.486, compared to measurements of 2.356 and 2.349 from 
the mispointed spectra.} 
% cf notes 23 Nov 15

\section*{Appendix~B --- New pulsation periods from $K$-band
photometry \label{s.appb}} % Sec. App. B

% K light curves of some very red sources

\begin{figure} % Fig. 19
\begin{center}
\includegraphics[width=3.4in]{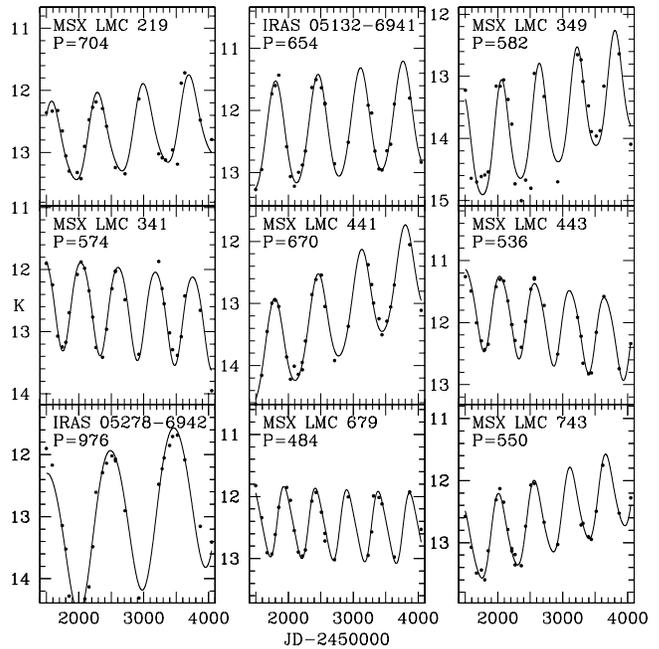} % klightcurves2.eps
\caption{The light curves at $K$ for the nine objects with 
newly reported pulsation periods.  The solid points are the
measured $K$ values and the lines are fits to the light 
curves using the periods displayed in the individual panels.  
The range on the $K$ axis is set to 3.2 magnitudes for all 
objects so that the differences in amplitudes are readily 
apparent.\label{f.kcurve}}
\end{center}
\end{figure}

\begin{deluxetable}{lrr} % Table 9 - from lightcurves/pw_nir_phot.dat
\tablecolumns{3}
\tablewidth{0pt}
\tablenum{9}
\tablecaption{Multi-epoch K-band photometry}
\label{t.kphot}
\tablehead{ \colhead{Target} & \colhead{JD$-$2450000} & \colhead{$K$ (mag.)} }
\startdata
MSX LMC 219     & 1501 & 12.36 \\
                & 1591 & 12.34 \\
                & 1675 & 12.32 \\
                & 1746 & 12.65 \\
                & 1800 & 13.05 \\
                & 1852 & 13.30 \\
                & 1976 & 13.32 \\
                & 2035 & 13.42 \\
                & 2090 & 12.90 \\
                & 2158 & 12.47 \\
\enddata
\tablecomments{Table~\ref{t.kphot} is published in its 
entirety in the electronic edition of the Astrophysical 
Journal.  A portion is shown here for guidance regarding its 
form and content.}
\end{deluxetable}

Some of the targets for Spitzer program 3505 (PI P.\ Wood) 
were very red and had no light curves available in the MACHO 
or OGLE monitoring programs.\footnote{The MACHO project
searched for gravitational lensing events from massive 
compact halo objects in the direction of the Magellanic 
Clouds.  \cite{fra05} show how this monitoring facilitated 
the study of long-period variables.  The OGLE project has
operated similarly.}  P.\ Wood led a long-term program to 
monitor many embedded evolved stars in the Magellanic Clouds 
using CASPIR, the Cryogenic Array Spectrometer Imager 
\citep{mcg94}, in the near-IR on the 2.3m telescope of the 
Australian National University at Siding Spring Observatory.  
\cite{gro07} describe the observational and data reduction 
procedures, and \cite{kam10} give the $K$- and $L$-band light 
curves of NGC~419 IR~1, NGC~419 MIR~1, NGC~1978 IR~1, and 
NGC~1978 MIR~1.  Figure~\ref{f.kcurve} presents the light 
curves at $K$ for nine additional objects.  The fitted light 
curves in the figure were made using a Fourier series of a 
single frequency and its first harmonic together with a term 
allowing a linear variation of $K$ with time.

These objects clearly show large pulsation amplitudes of 1.1 
to 2.5 magnitudes in $K$ as well as long-term variations in 
$K$ of up to 1 magnitude.  The latter variations are short 
segments of long-term variations \citep[e.g.][]{whi03} that 
are probably due to the episodic nature of dust-ejection 
mechanisms \citep[e.g.][]{win94}.  The large variations 
evident in the $K$ light curves highlight the fact that 
variability will introduce a significant scatter in 
quantities such as $M_{\rm bol}$ when it is computed from 
SEDs constructed from single-epoch photometry taken in
different bands at different times.  Similarly, colors 
computed using magnitudes obtained at different times for the 
two bands involved will also show a significant scatter due 
to variability.

Table~\ref{t.sample} identifies the nine new periods 
determined from $K$-band monitoring with the reference
``App.~B.''  Table~\ref{t.kphot} gives the multi-epoch 
$K$-band photometry.

\section*{Appendix~C --- New pulsation periods from
3--5~\mum\ photometry} % Sec. App. C

\begin{figure} % Fig. 20
\includegraphics[width=3.4in]{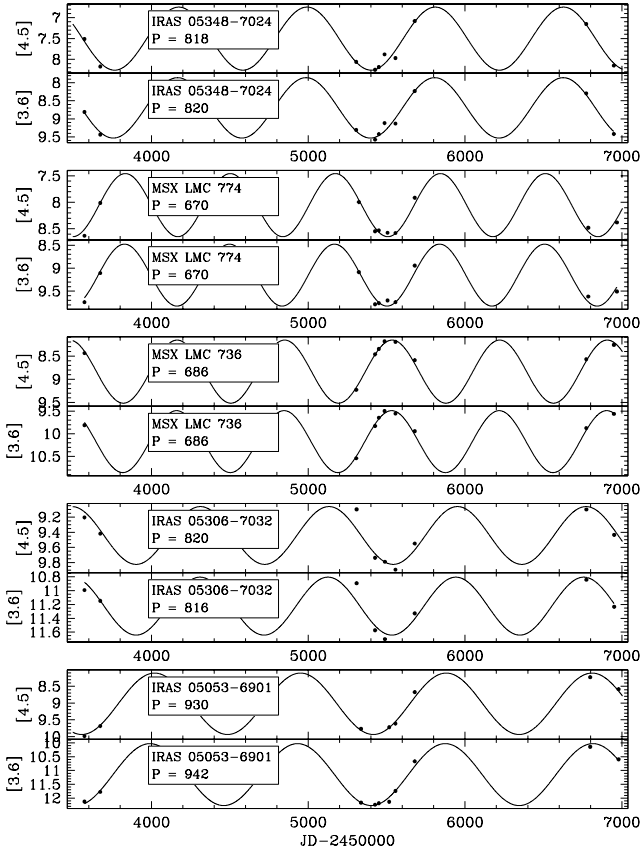} % mircurves.eps
\caption{The light curves at 3.6 and 4.5~\mum\ of five objects 
with previously unknown periods.  The solid points are the 
measured values and the lines are sine functions fitted to 
the light curves using the periods displayed in the panels.
\label{f.mircurves}}
\end{figure}

\begin{deluxetable*}{lrlcrrl} % Table 10 - from smc/lmc.cemir.tbl
\tablecolumns{7}
\tablewidth{0pt}
\tablenum{10}
\tablecaption{Multi-epoch 3--5~\mum\ photometry}
\label{t.mirphot}
\tablehead{ 
  \colhead{ } & \colhead{RA} & \colhead{Decl.}   & \colhead{ } & 
  \colhead{ } & \colhead{ } & \colhead{Data} \\
  \colhead{Target} & \multicolumn{2}{c}{(J2000)} & \colhead{MJD} & 
  \colhead{[3.6]} & \colhead{[4.5]} & \colhead{source\tablenotemark{a}}
}
\startdata
GM 780             &   8.905055 & $-$73.165641 & 54633 &  9.115 $\pm$ 0.059   &  8.609 $\pm$ 0.048   & IRAC         \\
                   &   8.905572 & $-$73.165577 & 54728 &  9.180 $\pm$ 0.034   &  8.686 $\pm$ 0.047   & IRAC         \\
                   &   8.905168 & $-$73.165589 & 55316 &  9.132 $\pm$ 0.025   &  8.672 $\pm$ 0.020   & ({\it WISE}) \\
                   &   8.905168 & $-$73.165589 & 55498 &  8.465 $\pm$ 0.035   &  8.040 $\pm$ 0.038   & ({\it WISE}) \\
                   &   8.905188 & $-$73.165607 & 56781 &  8.171 $\pm$ 0.019   &  7.764 $\pm$ 0.018   & ({\it WISE}) \\
                   &   8.905212 & $-$73.165612 & 56961 &  8.364 $\pm$ 0.025   &  7.959 $\pm$ 0.017   & ({\it WISE}) \\
MSX SMC 091        &   9.236177 & $-$72.421510 & 54633 &  9.293 $\pm$ 0.046   &  8.721 $\pm$ 0.036   & IRAC         \\
                   &   9.236628 & $-$72.421501 & 54728 &  9.328 $\pm$ 0.060   &  8.752 $\pm$ 0.053   & IRAC         \\
                   &   9.236252 & $-$72.421493 & 55318 & 10.131 $\pm$ 0.021   &  9.452 $\pm$ 0.036   & ({\it WISE}) \\
                   &   9.236252 & $-$72.421493 & 55499 &  9.406 $\pm$ 0.024   &  8.782 $\pm$ 0.056   & ({\it WISE}) \\
\enddata
\tablenotetext{a}{{\it WISE} data are corrected as described in 
Appendix~A.}
\tablecomments{Table~\ref{t.mirphot} is published in its 
entirety in the electronic edition of the Astrophysical 
Journal.  A portion is shown here for guidance regarding its 
form and content.}
\end{deluxetable*}

All the objects in this study have been observed multiple 
times at 3.6 and 4.5~\mum\ by {\it Spitzer} and at 3.4 and 
4.6~\mum\ by {\it WISE} (see Section~\ref{s.photo}).  We
examined if sufficient epochs were available from the two
spacecraft to generate mid-IR light curves and from these
determine pulsation periods.

In order to maximize the number of points in each light 
curve, we converted the data at 3.4 and 4.6~\mum\ from
{\it WISE} to the {\it Spitzer} wavelengths using the 
transformations from Appendix~A.  Table~\ref{t.mirphot}
presents the multi-epoch photometry at 3.6 and 4.5~\mum,
including the converted {\it WISE} photometry, for all of 
the Magellanic carbon stars in our sample.

A Fourier fit consisting of a single frequency was made to 
the resulting light curves at 3.6 and 4.5~\mum.  We found 
that reliable results were only possible for light curves 
with at least eight individual epochs and a range in the
photometry exceeding $\sim$0.5 magnitudes.  In addition, 
the periods determined independently at 3.6 and 4.5~\mum\ had 
to agree to better than 5\%.  

With the above constraints, we were able to fit the light 
curves of 42 objects.  Of these objects, 37 had periods 
previously determined from more extensive light curves 
available at shorter wavelengths, such as those from OGLE or
those discussed in Appendix~B.  Comparing the average of the 
periods from 3.6 and 4.5~\mum\ with the previously known 
periods showed that 31 of the 37 periods agreed to better 
than 5\%, a success rate of 84\%.

In our sample of fitted periods we found five sources with 
previously unknown periods.  Figure~\ref{f.mircurves} plots 
their light curves.  The averages of the periods from 3.6 and
4.5~\mum\ appear in Table~\ref{t.sample}, with the reference
for the periods given as ``App.~C.''  Based on our comparison 
of periods in the literature for the 37 other sources, we 
would expect that at least four of these five new periods are 
reliable to better than 5\%.

\acknowledgements

We thank the anonymous referee for helpful and constructive 
comments.
G.~C.~S.\ was supported by NASA through Contract Number 1257184 
issued by the Jet Propulsion Laboratory, California Institute of 
Technology under NASA contract 1407 and the NSF through Award 
1108645.  F.~K.\ received support from the Ministry of Science 
and Technology (MoST) of Taiwan, grant MOST104-2628-M-001-004-MY3.
This research relied on the following resources:
NASA's Astrophysics Data System, the Infrared Science Archive
at the Infrared Processing and Analysis Center, operated
by JPL, and the Simbad and VizieR databases, operated at the
Centre de Donn\'{e}es astronomiques de Strasbourg.


\begin{thebibliography}{}

\bibitem[Bernard-Salas et al.(2006)]{jbs06} Bernard-Salas, J., Peeters, E., 
  Sloan, G.C., et al.\ 2006, \apjl, 652, L29 % ok
\bibitem[Bernard-Salas et al.(2009)]{jbs09} Bernard-Salas, J., Peeters, E., 
  Sloan, G.C., et al.\ 2009, \apj, 699, 1541 % ok
\bibitem[Blum et al.(2006)]{blu06} Blum, R.~D., Mould, J.~R., Olsen, K.~A.,
  et al.\ 2006, \aj, 132, 2034 % ok
\bibitem[Bolatto et al.(2007)]{bol07} Bolatto, A.~D., Simon, J.~D.,
  Stanimirovi\'{c}, S., et al.\ 2007, \apj, 655, 212 % ok
\bibitem[Boyer et al.(2012)]{boy12} Boyer, M.~L., Srinivasan, S., Riebel,
  D., et al.\ 2012, \apj, 748, 40 % ok
\bibitem[Buchanan et al.(2006)]{buc06} Buchanan, C.~L., Kastner, J.~H.,
  Forrest, W.~J., et al.\ 2006, \aj, 132, 1890 % ok
\bibitem[Cioni et al.(2000)]{cio00} Cioni, M.-R., Loup, C., Habing, H.~J., 
  et al.\ 2000, \aaps, 144, 235 % ok
\bibitem[Cutri et al.(2006)]{cut06} Cutri, R.~M., Skrutskie, M.~F., van Dyk,
  S., et al.\ 2006, 2MASS 6X Point Source Working Database/Catalog, VizieR
  Catalog II/281 % ok
\bibitem[de Graauw et al.(1996)]{deg96} de Graauw, T., Haser, L.~N., 
  Beintema, D.~A., et al.\ 1996, \aap, 315, L49 % ok
\bibitem[Dell'Agli et al.(2015)]{del15} Dell'Agli, F., Garc\'{a}-Hern\'{a}ndez,
  D.~A., Ventura, P., et al.\ 2015, \mnras, 454, 4235 % ok
\bibitem[Feast(2013)]{fea13} Feast, M.~W.\ 2013, in Planets, Stars and Stellar
  Systems Vol. 5, Galactic Structure and Stellar Populations, ed.\ T.~D.\
  Oswalt \& G.\ Gilmore, 829 (Dordrecht:  Springer) % ok
\bibitem[Fraser et al.(2005)]{fra05} Fraser, O.~J., Hawley, S.~L., Cook, K.~H.,
  \& Keller, S.~C.\ 2005, \aj, 129, 768 % ok
\bibitem[Gardiner \& Hawkins(1991)]{gar91} Gardiner, L.~T., \& Hawkins,
  M.~R.~S.\ 1991, \mnras, 251, 174 % ok
\bibitem[Girardi et al.(2009)]{gir09} Girardi, L., Rubele, S., \& Kerber, L.\
  2009, \mnras, 394, L74 % ok
\bibitem[Glatt et al.(2008)]{gla08} Glatt, K., Grebel, E.~K., Sabbi, E., et 
  al.\ 2008, \aj, 136, 1703 % ok
\bibitem[Goebel \& Moseley(1985)]{gm85} Goebel, J.~H., \& Moseley, S.~H.\
  1985, \apj, 290, L35 % ok
\bibitem[Gordon et al.(2011)]{gor11} Gordon, K.~D., Meixner, M., Meade, M.~R., 
  et al.\ 2011, \aj, 142, 102 % ok
\bibitem[Groenewegen et al.(2009)]{gro09} Groenwegen, M.~A.~T., Sloan, G.~C., 
  Soszy\'{n}ski, I., \& Petersen, E.~A.\ 2009, \aap, 506, 1277 % ok
\bibitem[Groenewegen et al.(2007)]{gro07} Groenwegen, M.~A.~T., Wood, P.~R., 
  Sloan, G.~C., et al.\ 2007, \mnras, 376, 313 % ok
\bibitem[Gruendl et al.(2008)]{gru08} Gruendl, R.~A., Chu, Y.-H., Seale, J.~P.,
  et al.\ 2008, \apjl, 688, L9 % ok
\bibitem[Gullieuszik et al.(2012)]{gul12} Gullieuszik, M., Groenwegen, 
  M.~A.~T., Cioni, M.-R.~L., et al.\ 2012, \aap, 537, 105 % ok
\bibitem[Houck et al.(2004)]{hou04} Houck, J.~R., Roellig, T.~L., \& van
  Cleve, J., et al.\ 2004, \apjs, 154, 18 % ok
\bibitem[Jones et al.(1977)]{jon78} Jones, B., Merrill, K.~M., Puetter, R.~C.,
  \& Willner, S.~P.\ 1977, \aj, 83, 1437 % ok
\bibitem[Jones(2015)]{dj15} Jones, D.\ 2015, in The Physics of Evolved Stars: 
  A Conference Dedicated to the Memory of Olivier Chesneau, ed.\ E.\ Lagadec, 
  F.\ Millour \& T.\ Lanz, EAS Publications Series, 71–72, 113 % ok
\bibitem[Jones et al.(2015)]{jon15} Jones, O.~C. Meixner, M., Sargent, B.~A.,
  et al.\ 2015, \apj, 811, 145 % ok
\bibitem[Kamath et al.(2010)]{kam10} Kamath, D., Wood, P.~R., Soszy\'{n}ski, 
  I., \& Lebzelter, T.\ 2010, \mnras, 408, 522 % ok
\bibitem[Kato et al.(2007)]{kat07} Kato, D., Nagashima, Ch., Nagayama, T.,
  et al.\ 2007, \pasj, 59, 615 % ok
\bibitem[Kemper et al.(2010)]{kem10} Kemper, F., Woods, P.~M., Antoniou, V., 
  et al. 2010, \pasp, 122, 683 % ok
\bibitem[Kervella et al.(2015)]{ker15} Kervella, P., Montarg\`{e}s, M., 
  Lagadec, E., et al.\ 2015, \aap, 578, 77 % ok
\bibitem[Kervella et al.(2014)]{ker14} Kervella, P., Montarg\`{e}s, M., 
  Ridgway, S.~T., et al.\ 2014, \aap 564 88 % ok
\bibitem[Kessler et al.(1996)]{kes96} Kessler, M.~F., Steinz, J.~A., 
  Anderegg, M.~E., et al.\ 1996, \aap, 315, L27 % ok
\bibitem[Kraemer et al.(2006)]{kra06} Kraemer, K.~E., Sloan, G.~C., 
  Bernard-Salas, J., et al.\ 2006, \apjl, 652, L25 % ok
\bibitem[Kraemer et al.(2002)]{kra02} Kraemer, K.~E., Sloan, G.~C., Price,
  S.~D., \& Walker, H.~J.\ 2002, \apjs, 140, 389 % ok
\bibitem[Kraemer et al.(2005)]{kra05} Kraemer, K.~E., Sloan, G.~C., Wood,
  P.~R., et al.\ 2005, \apjl, 631, L147 % ok
\bibitem[Lagadec \& Chesneau(2014)]{lc14} Lagadec, E., \& Chesneau, O.\ 2014,
  in Why Galaxies Care about AGB Stars III: A Closer Look in Space and Time, 
  ed.\ F.\ Kerschbaum, R.~F.\ Wing, and J.\ Hron, ASPC, 497, 145 (San
  Francisco, ASP) % ok
\bibitem[Lagadec \& Zijlstra(2008)]{lz08} Lagadec, E., \& Zijlstra, A.~A.\
  2008, \mnras, 390, L59 % ok
\bibitem[Lagadec et al.(2010)]{lag10} Lagadec, E., Zijlstra, A.~A., Mauron, N.,
  et al.\ 2010, \mnras, 403, 1331 % ok
\bibitem[Lagadec et al.(2007)]{lag07} Lagadec, E., Zijlstra, A.~A., Sloan,
  G.~C., et al.\ 2007, \mnras, 376, 1270 % ok
\bibitem[Lebofsky \& Rieke(1977)]{lr77} Lebofsky, M.~J., \& Rieke, G.~H.\
  1977, \aj, 82, 646 % ok
\bibitem[Lebouteiller et al.(2010)]{leb10} Lebouteiller, V., Bernard-Salas, J.,
  Sloan, G.~C., \& Barry, D.~J.\ 2010, \pasp, 122, 188 % ok
\bibitem[Leisenring et al.(2008)]{lei08} Leisenring, J.~M., Kemper, F., \&
  Sloan, G.~C.\ 2008, \apj, 681, 1557 % ok
\bibitem[Lombaert et al.(2012)]{lom12} Lombaert, R., de Vries, B.~L., de
  Koter, A., et al.\ 2012, \aap, 544, L18 % ok
\bibitem[Maercker, et al.(2012)]{mae12} Maercker, M., Mohamed, S., Vlemmings, 
  W.~H.~T., et al.\ 2012, Nature, 490, 232 % ok
\bibitem[Mainzer et al.(2014)]{mai14} Mainzer, A., Bauer, J., Cutri, R.~M.,
  et al.\ 2014, \apj, 792, 30 % ok
\bibitem[Matsuura et al.(2009)]{mat09} Matsuura, M., Barlow, M.~J., Zijlstra,
  A.~A., et al.\ 2009, \mnras, 396, 918 % ok
\bibitem[Matsuura et al.(2014)]{mat14} Matsuura, M., Bernard-Salas, J.,
  Lloyd Evans, T., et al.\ 2014, \mnras, 439, 1472 % ok
\bibitem[Matsuura et al.(2006)]{mat06} Matsuura, M., Wood, P.~R., Sloan, G.~C.,
  et al.\ 2006, \mnras, 371, 415 % ok
\bibitem[Matsuura et al.(2005)]{mat05} Matsuura, M., Zijlstra, A.~A., van
  Loon, J.~Th., et al.\ \aap, 434, 691 % ok
\bibitem[McGregor et al.(1994)]{mcg94} McGregor, P.~J., Hart, J.,
  Hoadley, D., \& Bloxham, G. 1994, in Infrared Astronomy with Arrays,
  ed.\ I McLean (Dordrecht:  Kluwer), 299 % ok
\bibitem[Meixner et al.(2006)]{mei06} Meixner, M., Gordon, K.~D., Indebetouw,
  R., et al.\ 2006, \aj, 132, 2268 % ok
\bibitem[Nidever et al.(2013)]{nid13} Nidever, D.~L., Monachesi, A., Bell, 
  E.~F., et al.\ 2013, \apj, 779, 145 % ok
\bibitem[Nishida et al.(2000)]{nis00} Nishida, S., Tanab\'{e}, T., Nakada, Y.,
  et al.\ 2000, \mnras, 313, 136 % ok
\bibitem[Pietrzy\'{n}ski et al.(2013)]{pie13} Pietrzy\'{n}ski, G., Graczyk, D.,
  Gieren, W., et al.\ 2013, Nature, 495, 76 % ok
\bibitem[Piatti(2012)]{pia12} Piatti, A.~E.\ 2012, \mnras, 422, 1109 % ok
\bibitem[Piatti(2013)]{pia13} Piatti, A.~E., \& Giesler, D.\ 2013, \apj, 145, 
  17 % ok
\bibitem[Price et al.(2001)]{pri01} Price, S.~D., Egan, M.~P., Carey, S.~J., 
  Mizuno, D.~R., \& Kuchar, T.~A.\ 2001, \aj, 121, 2819 % ok
\bibitem[Raimondo et al.(2005)]{rai05} Raimondo, G., Cioni, M.-R.~L., Rejkuba, 
  M., Silva, D.~R.\ 2005, \aap, 438, 521 % ok
\bibitem[Riebel et al.(2015)]{rie15} Riebel, D., Boyer, M.~L., Srinivasan, S.,
  et al.\ 2015, \apj, 807, 1 % ok
\bibitem[Riebel et al.(2012)]{rie12} Riebel, D., Srinivasan, S., Sargent, B.,
  \& Meixner, M.\ 2015, \apj, 753, 71 % ok
\bibitem[Rubele et al.(2015)]{rub15} Rubele, S., Girardi, L., Kerber, L., et 
  al.\ 2015, \mnras, 449, 639 % ok
\bibitem[Skrutskie et al.(2006)]{skr06} Skrutskie, M.~F., Cutri, R.~M.,
  Stiening, R., et al.\ 2006, \aj, 131, 1163 % ok
\bibitem[Sloan \& Egan(1995)]{slo95} Sloan, G.C., \& Egan, M.~P.\ 1995, \apj,
  444, 452 % ok
\bibitem[Sloan et al.(2015a)]{slo15a} Sloan, G.~C., Herter, T.~L., 
  Charmandaris, V., et al.\ 2015a, \aj, 149, 11 % ok
\bibitem[Sloan et al.(2006)]{slo06} Sloan, G.~C., Kraemer, K.~E., Matsuura,
  M., et al.\ 2006, \apj, 645, 1118 % ok
\bibitem[Sloan et al.(2008)]{slo08} Sloan, G.~C., Kraemer, K.~E., Wood, P.~R., 
  et al.\ 2008, \apj, 686, 1056 % ok
\bibitem[Sloan et al.(2015b)]{slo15b} Sloan, G.~C., Lagadec, E., Kraemer, 
  K.~E., et al.\ 2015, in Why Galaxies Care about AGB Stars III, ed.\ F.\ 
  Kerschbaum, J.\ Hron, \& R.\ Wing, ASP Conf.\ Series, 497, 429 % ok
\bibitem[Sloan et al.(2014)]{slo14} Sloan, G.~C., Lagadec, E., Zijlstra, 
  A.~A., et al.\ 2014, \apj, 791, 28 % ok
\bibitem[Sloan et al.(2012)]{slo12} Sloan, G.~C., Matsuura, M., Lagadec, E., 
  et al.\ 2012, \apj, 752, 140 % ok
\bibitem[Sloan \& Price(1995)]{sp95} Sloan, G.~C., \& Price, S.~D.\ 1995,
  \apj, 451, 758 % ok
\bibitem[Soker \& Livio(1989)]{sl89} Soker, N., \& Livio, M.\ 1989, \apj, 339,
  268 % ok
\bibitem[Soszy\'{n}ski et al.(2009)]{sos09} Soszy\'{n}ski, Udalski, A.,
  Szyma\'{n}ski, M.~K., et al.\ 2009, AcA, 59, 335 % ok
\bibitem[Soszy\'{n}ski et al.(2011)]{sos11} Soszy\'{n}ski, Udalski, A.,
  Szyma\'{n}ski, M.~K., et al.\ 2011, AcA, 61, 217 % ok
\bibitem[Srinivasan et al.(2011)]{sri11} Srinivasan, S., Sargent, B.~A., \&
  Meixner, M.\ 2011, \aap, 532, 54 % ok
\bibitem[Srinivasan et al.(2016)]{sri16} Srinivasan, S., Boyer, M.~L., 
  Kemper, F., et al.\ 2016, \mnras, 457, 2814  % ok
\bibitem[Ulaczyk et al.(2013)]{ula13} Ulaczyk, K., Szyma\'{n}ski, M.~K., 
  Udalski, A., et al.\ 2013, AcA 63, 1 % ok
\bibitem[van Loon et al.(2008)]{vl08} van Loon, J.~Th., Cohen, M., Oliveira, 
  J.~M., et al.\ 2008, \aap, 487, 1055 % ok
\bibitem[van Loon et al.(2006)]{vl06} van Loon, J.~Th., Marshall, J.~R.,
  Cohen, M., et al.\ 2006, \aap, 447, 971 % ok
\bibitem[van Loon et al.(2005)]{vl05} van Loon, J.~Th., Marshall, J.~R.,
  \& Zijlstra, A.~A.\ 2005, \aap, 442, 597 % ok
\bibitem[Vassiliadis \& Wood(1993)]{vw93} Vassiliadis, E., \& Wood, P.~R.\
  1993, \apj, 413, 641 % ok
\bibitem[Ventura et al.(2016)0000000]{ven16} Ventura, P., Karakas, A.~I., 
 Dell'Agli, F., et al. \mnras, 457, 1456
\bibitem[Werner et al.(2004)]{wer04} Werner, M.~W., Roellig, T.~L., Low,
  F.~J., et al.\ 2004, \apjs, 154, 1 % ok
\bibitem[Whitelock et al.(1989)]{whi89} Whitelock P.~A., Feast M.~W., Menzies 
  J.~W., Catchpole R.~M.\ 1989, \mnras, 238, 769 % ok
\bibitem[Whitelock et al.(2006)]{whi06} Whitelock, P.~A., Feast, M.~W., 
  Marang, F., Groenewegen, M.~A.~T.\ 2006, \mnras, 369, 751 % ok
\bibitem[Whitelock et al.(2003)]{whi03} Whitelock, P.~A., Feast, M.~W.,
  van Loon, J.~Th., \& Zijlstra, A.~A.\ 2003, \mnras, 342, 86 % ok
\bibitem[Winters et al.(1994)]{win94} Winters, J.~M., Fleischer, A.~J.,
  Gauger, A., \& Sedlmayr, E.\ 1994, \aap, 290, 623 % ok
\bibitem[Woitke(2006)]{woi06} Woitke, P.\ 2006, \aap, 452, 537 % ok
\bibitem[Wright et al.(2010)]{wri10} Wright, E.~L., Eisenhardt, P.~R.~M., 
  Mainzer, A.~K., et al.\ 2010, \aj, 140, 1868 % ok
\bibitem[Zaritsky et al.(2002)]{zar02} Zaritsky, D., Harris, J., Thompson,
  I.~B., Grebel, E.~K., \& Massey, P.\ 2002, \aj, 123, 855 % ok
\bibitem[Zaritsky et al.(2004)]{zar04} Zaritsky, D., Harris, J., Thompson,
  I.~B., \& Grebel, E.~K.\ 2004, \aj, 128, 1606 % ok
\bibitem[Zhang et al.(2009)]{zha09} Zhang, K., Jiang, B.~W., \& Li, A.\ 2009,
  \apj, 702, 680 % ok
\bibitem[Zijlstra et al.(2006)]{zij06} Zijlstra, A.~A., Matsuura, M., Wood,
  P.~R., et al.\ 2006, \mnras, 370, 1961 % ok
% BIB repairs - comma even for 2 authors, if 6+, list 3 et al.

\end{thebibliography}
\end{document}